\documentclass[aps, pre, onecolumn, floatfix]{revtex4}
\usepackage{blindtext}
\usepackage{graphicx}
\usepackage{amssymb, amsfonts, amsmath, amsthm, mathtools}
\usepackage{epsf}
\usepackage{subfigure}
\usepackage{epstopdf}
\usepackage{upgreek}
\usepackage{txfonts, pxfonts}
\usepackage{graphics}
\usepackage[usenames, dvipsnames]{xcolor}
\usepackage{float}
\usepackage{mathrsfs}
\usepackage{chemarrow}
\usepackage{bigstrut}
\usepackage{array}
\usepackage[utf8]{inputenc}
\usepackage[T1]{fontenc}
\usepackage{lmodern}

\newcommand{\be}{\begin{equation}}
\newcommand{\ee}{\end{equation}}
\newcommand{\bea}{\begin{eqnarray}}
\newcommand{\eea}{\end{eqnarray}}

\begin{document}
\title{Counting statistics and microreversibility in stochastic models of transistors}
\author{Jiayin Gu}
\email{jiaygu@ulb.ac.be}
\author{Pierre Gaspard}
\email{gaspard@ulb.ac.be}
\affiliation{Center for Nonlinear Phenomena and Complex Systems, Universit\'e Libre de Bruxelles (U.L.B.), Code Postal 231, Campus Plaine, B-1050 Brussels, Belgium}

\begin{abstract}
Multivariate fluctuation relations are established in three stochastic models of transistors, which are electronic devices with three ports and thus two coupled currents.  In the first model, the transistor has no internal state variable and particle exchanges between the ports is described as a Markov jump process with constant rates.  In the second model, the rates linearly depend on an internal random variable, representing the occupancy of the transistor by charge carriers.  The third model has rates nonlinearly depending on the internal occupancy.  For the first and second models, finite-time multivariate fluctuation relations are also established giving insight into the convergence towards the asymptotic form of multivariate fluctuation relations in the long-time limit.  For all the three models, the transport properties are shown to satisfy Onsager's reciprocal relations in the linear regime close to equilibrium as well as their generalizations holding in the nonlinear regimes farther away from equilibrium, as a consequence of microreversibility.
\end{abstract}

\maketitle

\section{Introduction}

Microreversibility is a fundamental symmetry of Nature, which manifests itself at the microscale in the movements and electromagnetic interactions of particles composing matter.  As shown by Onsager in 1931~\cite{O31a,O31b}, microreversibility also manifests itself at larger scales in transport phenomena running in linear regimes close to thermodynamic equilibrium.  In the nonlinear regimes farther away from equilibrium, energy dissipation and irreversible entropy production become dominant effects and we may wonder if there remain some signatures of underlying microreversibility.  Theoretical studies predict that such signatures exist in the form of relations generalizing Onsager's reciprocal relations to the {\it nonlinear} response coefficients of the mean currents flowing across nonequilibrium systems and their statistical cumulants at arbitrarily high orders in the deviations from equilibrium \cite{BK77,BK79,BK81,S92,AG04,AG06JSM,AG07JSM,AGMT09,SU08,HPPG11,G13NJP,BG18,BG19}.  These predictions that are the consequences of microreversibility have not yet been tested experimentally.

Recently, the authors of the present paper have shown that these predictions should be observed in the nonlinear transport properties of transistors at room temperature \cite{GG19}.  These electronic devices are coupling together two electric currents, so that their characteristics should obey not only Onsager's reciprocal relations in the close vicinity of equilibrium (corresponding to zero applied voltage), but also their generalizations in the strongly nonlinear regimes where transistors are known to function when voltages are applied.  In our previous work \cite{GG19}, transistors have been described as spatially extended systems where the charge carriers undergo diffusion-reaction stochastic processes including their Coulomb interactions and obeying local detailed balance in consistency with microreversibility.  The study reported in Ref.~\cite{GG19} has shown that the Onsager reciprocal relations and their generalizations to the nonlinear transport properties are indeed satisfied.  Moreover, we have shown that transistors have currents satisfying the so-called multivariate fluctuation relations (i.e., fluctuation relations for multiple currents) \cite{AG07JSP,EHM09, CHT11,S12}.  These fluctuation relations are the consequences of microreversibility for the full counting statistics of the currents that are coupled together in transistors.  Fluctuation relations are known to hold asymptotically in the long-time limit and an important issue is to determine the time scale of convergence towards the asymptotic behavior.  In this regard, we should mention that finite-time fluctuation relations have been obtained for particular stochastic systems crossed by a single current \cite{HS07JSM,AG08PRE,GK18JSM,GGHK18JSM}, allowing us to investigate in detail the issue of convergence in such cases.

The purpose of the present paper is to study and compare three different Markovian stochastic processes modelling transistors in order to understand whether they obey multivariate fluctuation relations at finite times or asymptotically at long enough times, and what are the implications of such relations.  Transistors are here considered as compact systems without internal random variable or a single one.   Indeed, because of their smallness, transistors often behave as a whole with their internal modes essentially driven by the voltage sources, as a consequence of the fast diffusion of the charge carriers and Coulomb interaction between them.

The first stochastic model we consider has no internal variable, the random currents across the transistor being described as a Markovian process of particle exchanges at constant rates between the three ports  connecting the device to three reservoirs at different voltages.  In the second and third models, the transistor is supposed to have a single internal random variable, representing its occupancy by charge carriers.  The transition rates linearly depend on this internal variable in the second model, but nonlinearly in the third model.  This latter is a model of conductive island or quantum dot in contact with the reservoirs through mesoscopic tunnel junctions \cite{AWBMJ91}.  Such a model describes, in particular, single-electron transistors \cite{AL86,K92}.

Our goal is to establish the linear and nonlinear transport properties of these models of transistors, in order to determine whether they obey finite-time or asymptotic fluctuation relations and if the consequences of microreversibility are satisfied in these models.

The plan of this paper is the following. Generalities about the multivariate current fluctuation relation and its implications are presented in Section~\ref{Gen}.  In Section~\ref{Model1}, we investigate the transport properties of the model with constant rates.  In Section~\ref{Model2}, the model of transistor with rates linearly depending on an internal variable is studied.  In Section~\ref{Model3}, the model with rates nonlinearly depending on an internal variable is analyzed.  The conclusion and perspectives are given in Section~\ref{Conclusion}.

\section{Generalities}
\label{Gen}

\subsection{Multivariate fluctuation relation for currents}

\begin{figure}[h]
\begin{minipage}[t]{0.32\hsize}
\resizebox{1.0\hsize}{!}{\includegraphics{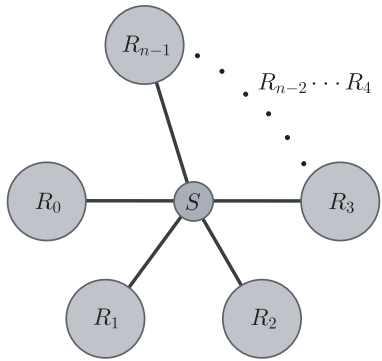}}
\end{minipage}
\caption{Schematic representation of a system $S$ in contact with $n$ particle reservoirs $R_0,R_1,R_2,\dots,R_{n-1}$.}
\label{fig_model}
\end{figure}

Let us consider a system $S$ exchanging particles with $n$ reservoirs $R_i$ ($i=0,1,\dots,n-1$), as shown in Fig.~\ref{fig_model}.  The whole system is supposed to be isothermal at the temperature $T$.  Besides, the thermodynamic state of each reservoir is characterized by its chemical potential $\mu_i$ ($i=0,1,\dots,n-1$).  If the chemical potentials take different values, the whole system is out of equilibrium and will reach a nonequilibrium steady state after long enough transients.   The control parameters of this nonequilibrium state are the thermodynamic forces also called affinities \cite{D36,P67,GM84,C85}.  There are as many independent affinities as there are differences between chemical potentials.  Taking $R_0$ as the reference reservoir, the affinities are defined as
\be\label{Aff-dfn}
A_i \equiv \beta(\mu_i-\mu_0) \, ,
\ee
where $\beta=(k_{\rm B}T)^{-1}$ is the inverse temperature, $k_{\rm B}$ being Boltzmann's constant.
These affinities are collectively denoted by the vector ${\bf A}=(A_1,\dots,A_i,\dots,A_{n-1})$.
At thermodynamic equilibrium, all the chemical potentials take the same value, so that all the affinities are vanishing, ${\bf A}={\bf 0}$.

At the mesoscopic level of description, the exchange of particles between the system and the reservoirs is assumed to be ruled by a Markovian stochastic process.  This process rules the time evolution of the random numbers of particles $Z_i$ ($i=1,\dots,n-1$) exchanged from the reservoir $R_i$ ($i=1,\dots,n-1$) to the reference reservoir $R_0$, as well as on possible random variables that are internal to the system $S$, such as the occupancy of some internal quantum levels.

After the relaxation of the system towards the steady state of affinities $\bf A$, the process is characterized by the joint probability distribution $P({\bf Z},t;{\bf A})$ to have observed the particle exchanges ${\bf Z}=(Z_1,\dots,Z_i,\dots,Z_{n-1})$ during the time interval $[0,t]$.  The knowledge of this joint probability distribution and all its cumulants determine the so-called full counting statistics.   

As a consequence of microreversibility, the joint probability distribution obeys the multivariate fluctuation relation \cite{G13NJP},
\be
\frac{P({\bf Z},t;{\bf A})}{P(-{\bf Z},t;{\bf A})}\sim_{t\to\infty}\exp({\bf A}\cdot{\bf Z}) \, ,
\label{eq_fluctuation_relation}
\ee
meaning that the ratio of opposite fluctuations of particle exchanges goes exponentially in time under nonequilibrium conditions ${\bf A}\ne{\bf 0}$.  The fluctuation relation is valid arbitrarily far away from equilibrium.  At thermodynamic equilibrium where ${\bf A}={\bf 0}$, we recover the principle of detailed balance according to which the probabilities of opposite fluctuations are equiprobable.  If random currents are defined over the finite time interval $[0,t]$ according to $\pmb{\cal J}\equiv{\bf Z}/t$, we see that Eq.~(\ref{eq_fluctuation_relation}) is the fluctuation relation for all the currents flowing across the system.   As a consequence of the multivariate fluctuation relation~(\ref{eq_fluctuation_relation}), the cumulant generating function
\be
Q(\pmb{\lambda};{\bf A}) \equiv \lim_{t\to\infty} -\frac{1}{t} \, \ln \sum_{\bf Z} P({\bf Z},t;{\bf A}) \, {\rm e}^{-\pmb{\lambda}\cdot{\bf Z}} \, .
\ee
obeys the following symmetry relation
\be\label{FR_Q}
Q(\pmb{\lambda};{\bf A}) = Q({\bf A}-\pmb{\lambda};{\bf A}) \, ,
\ee
which can be proved using a time-evolution operator that is modified to include the parameters~$\pmb{\lambda}$ counting particle exchanges \cite{AG06JSM,AG07JSP,LS99,LM09}.

In general, the multivariate fluctuation relation holds asymptotically in time for $t\to\infty$.  However, it has been shown for specific processes between two reservoirs that the fluctuation relation may hold at every time with respect to some time-dependent affinity \cite{AG08PRE,GK18JSM,GGHK18JSM}.

\subsection{Implications for linear and nonlinear response coefficients}
\label{Gen-resp_coeff}

Now, the mean currents and their diffusivities can be obtained by taking the successive derivatives of the generating function with respect to the counting parameters:
\bea
&& J_{i}({\bf A})\equiv \lim_{t\to\infty} \frac{1}{t} \, \langle Z_i(t)\rangle_{\bf A} =\left.\frac{\partial Q(\pmb{\lambda};{\bf A})}{\partial\lambda_{i}}\right\vert_{\pmb{\lambda}={\bf 0}} , \label{J} \\
&& D_{ij}({\bf A})\equiv \lim_{t\to\infty} \frac{1}{2t} \, \langle[Z_i(t)-J_i\,t][Z_j(t)-J_j\,t]\rangle_{\bf A} =\left.-\frac{1}{2}\frac{\partial^2 Q(\pmb{\lambda};{\bf A})}{\partial\lambda_{i}\partial\lambda_{j}}\right\vert_{\pmb{\lambda}={\bf 0}} , \label{Diff}
\eea
where the notation $\langle\cdot\rangle$ denotes average over the sample data. By definition, the diffusivities satisfy the symmetry relation $D_{ij}=D_{ji}$.

Besides, the mean currents can be expanded in power series of the affinities as
\be\label{J-expanded}
J_{i}=\sum_{j}L_{i,j}A_{j}+\frac{1}{2}\sum_{j,k}M_{i,jk}A_{j}A_{k}+\cdots
\ee
in terms of the response coefficients defined by
\bea
&& L_{i,j}\equiv \left.\frac{\partial J_{i}}{\partial A_{j}}\right\vert_{{\bf A}={\bf 0}}=\left.\frac{\partial^2 Q(\pmb{\lambda};{\bf A})}{\partial\lambda_{i}\partial A_{j}}\right\vert_{\pmb{\lambda}={\bf A}={\bf 0}} , \label{L_ij}\\
&& M_{i,jk}\equiv \left.\frac{\partial^2 J_{i}}{\partial A_{j}\partial A_{k}}\right\vert_{{\bf A}={\bf 0}}=\left.\frac{\partial^3 Q(\pmb{\lambda};{\bf A})}{\partial\lambda_{i}\partial A_{j}\partial A_{k}}\right\vert_{\pmb{\lambda}={\bf A}={\bf 0}} .
\label{M_ijk}
\eea
We note that $M_{i,jk}=M_{i,kj}$ by the definition of these coefficients.

According to the symmetry relation of fluctuation theorem, we can derive to the so-called fluctuation-dissipation relations
\be
L_{i,j}=D_{ij}({\bf A}=0) \, , \label{eq_FD_relation}
\ee
where the diffusivities $D_{ij}({\bf A}=0)$ is given by Eq.~(\ref{Diff}).  The Onsager reciprocal relations immediately follows
\be
L_{i,j}=L_{j,i} \, . \label{eq_Onsager_reciprocal_relations}
\ee
We can also obtain relations for nonlinear response coefficients such that
\be
M_{i,jk}=R_{ij,k}+R_{ik,j} \, , \label{eq_nonlinear_response_relations}
\ee
where
\be\label{R_ijk}
R_{ij,k}\equiv \left.\frac{\partial D_{ij}({\bf A})}{\partial A_{k}}\right\vert_{{\bf A}={\bf 0}}=\left.-\frac{1}{2}\frac{\partial^3 Q(\pmb{\lambda};{\bf A})}{\partial\lambda_{i}\partial\lambda_{j}\partial A_{k}}\right\vert_{\pmb{\lambda}={\bf A}={\bf 0}} 
\ee
are the first responses of the diffusivities, which satisfy the identity $R_{ij,k}=R_{ji,k}$ because of their definition.

Further relations are known between the higher-order response coefficients and cumulants \cite{AG06JSM,AG07JSM,BG18,BG19}.

\subsection{Central limit theorem}

According to the central limit theorem holding in the long-time limit, the joint probability distribution is expected to become the Gaussian distribution
\be
{\mathscr P}({\bf Z},t)=\frac{1}{\sqrt{(4\pi t)^{n-1}\det{\boldsymbol{\mathsf D}}}}\exp\left[-\frac{1}{4t} ({\bf Z}-{\bf J}t)\cdot{\boldsymbol{\mathsf D}}^{-1}\cdot({\bf Z}-{\bf J}t)\right] , \label{eq_Gaussian_distribution}
\ee
in terms of the mean currents ${\bf J}=(J_i)_{i=1}^{n-1}$ and the diffusion matrix ${\boldsymbol{\mathsf D}}=(D_{ij})_{i,j=1}^{n-1}$ given by Eqs.~(\ref{J}) and~(\ref{Diff}).  These quantities are evaluated under nonequilibrium conditions, so that they both depend on the affinities: ${\bf J}={\bf J}({\bf A})$ and ${\boldsymbol{\mathsf D}}={\boldsymbol{\mathsf D}}({\bf A})$.

If the Gaussian distribution~(\ref{eq_Gaussian_distribution}) was substituted in the fluctuation relation~(\ref{eq_fluctuation_relation}), the affinities would be given by ${\bf A}={\boldsymbol{\mathsf D}}^{-1}\cdot{\bf J}$, so that the currents would depend on the affinities according to the relation: ${\bf J}({\bf A})={\boldsymbol{\mathsf D}}({\bf A})\cdot{\bf A}$.  However, the currents are known to also depend on the third and higher statistical cumulants, which play an important role in the nonlinear regimes~\cite{G13NJP}.  Including the third cumulants ${\boldsymbol{\mathsf C}}=(C_{ijk})$, the currents are actually given by
\be
{\bf J}({\bf A})={\boldsymbol{\mathsf D}}({\bf A})\cdot{\bf A}-\frac{1}{4}\, {\boldsymbol{\mathsf C}}({\bf A}):{\bf A}{\bf A}+O({\bf A}^3) \, ,
\ee
where the dot (respectively, the colon) denotes the contraction over one index (respectively, two indices)~\cite{G13NJP}.  Therefore, the affinities cannot be accurately obtained using the central limit theorem outside the linear regime.  The reason is that the central limit theorem only capture the top of the joint probability distribution $P({\bf Z},t;{\bf A})$ with its first and second cumulants~(\ref{J}) and~(\ref{Diff}), although the fluctuation relation~(\ref{eq_fluctuation_relation}) is a large-deviation property fully characterizing the distribution with all its cumulants. In particular, the cumulants higher than the first and second ones are essential to determine the tails of the distribution and thus the affinities.

In this regard, a fundamental issue is to understand the stochastic process beyond the central limit theorem.  This issue is addressed for three different models, for which the fluctuation relation and its implications will be investigated.

\subsection{Entropy production}

The entropy production rate, which is internal to the system \cite{P67,GM84,C85}, can be directly evaluated in steady states using the multivariate fluctuation relation~(\ref{eq_fluctuation_relation}) for the currents according to 
\be\label{entrprod}
\frac{1}{k_{\rm B}} \frac{{\rm d}_{\rm i}S}{{\rm d}t} = \lim_{t\to\infty} \frac{1}{t} \sum_{\bf Z} P({\bf Z},t;{\bf A}) \, \ln \frac{P({\bf Z},t;{\bf A})}{P(-{\bf Z},t;{\bf A})} = {\bf A}\cdot{\bf J}({\bf A}) \ge 0 \, ,
\ee
which is always non negative in accordance with the second law of thermodynamics.  

Since the mean electric currents are related by ${\bf I}=e{\bf J}$ to the mean particle currents $\bf J$, where $e=\pm\vert e\vert$ is the electric charge of the exchanged particles, and the affinities can be expressed as ${\bf A}=\beta e {\bf V}$ in terms of the voltages $\bf V$, the entropy production rate given by Eq.~(\ref{entrprod}) has the following equivalent expression,
\be\label{entrprod2}
\frac{{\rm d}_{\rm i}S}{{\rm d}t} = \frac{1}{T} \, {\bf V}\cdot{\bf I}({\bf V}) \ge 0 \, ,
\ee
which is the dissipated power divided by the temperature $T$, as required.

\section{Model with constant rates}
\label{Model1}

For a general $n$-reservoir system (see Fig. \ref{fig_model}), a coarse-grained model can be used to approximate the full description of particle transitions through the system.  In this coarse-grained model, two transition rates are hypothesized to exist between any two different reservoirs in the long-time limit, that is
\be
R_i\autorightleftharpoons{$\scriptstyle W_{ij}$}{$\scriptstyle W_{ji}$}R_j \hspace{1cm} i,j=0,1,\dots,n-1; \ i\neq j \, .
\ee

\subsection{Counting statistics} 
\label{Countstat_model1}

The counting statistics of particle transfers between the $n-1$ reservoirs ($i=1,\dots,n-1$) and the reference reservoir during some time interval $t$ is performed, and the numbers of particle transfers are given in vectorial notation by ${\bf Z}=(Z_1,\dots,Z_{n-1})$.  The time evolution of the probability distribution $P({\bf Z},t)$ is ruled by the master equation
\be
\frac{\rm d}{{\rm d}t}P({\bf Z},t) =\left\{\sum_{i=1}^{n-1}\left[\left({\rm e}^{-\partial_{Z_i}}-1\right)W_{i0}+\left({\rm e}^{+\partial_{Z_i}}-1\right)W_{0i}\right] + \sum_{i=1}^{n-1}\sum_{j=i+1}^{n-1}\left({\rm e}^{-\partial_{Z_i}}{\rm e}^{+\partial_{Z_j}}-1\right)W_{ij}\right\}P({\bf Z},t) \, , \label{eq_general_master_equation}
\ee
expressed in terms of the raising and lowering operators such that
\be
{\rm e}^{\pm\partial_{Z_i}} P(Z_1,\dots,Z_i,\dots,Z_{n-1},t) = P(Z_1,\dots,Z_i \pm 1,\dots,Z_{n-1},t) \, .
\ee

The finite generating function of the signed cumulated fluxes is defined as \cite{G04}
\be\label{G-fn}
G(s_1,\dots,s_{n-1},t)\equiv\sum_{Z_1,\dots,Z_{n-1}=-\infty}^{+\infty}s_1^{Z_1}\cdots s_{n-1}^{Z_{n-1}}\, P(Z_1,\dots,Z_{n-1},t) \, ,
\ee
whose evolution equation can be deduced from Eq. (\ref{eq_general_master_equation}), reading
\bea
&&\partial_tG(s_1,\dots,s_{n-1},t)=\Bigg[\sum_{i=1}^{n-1}\left(W_{i0}s_i+\frac{W_{0i}}{s_i}-W_{i0}-W_{0i}\right) \nonumber \\
&& +\sum_{i=1}^{n-1}\sum_{j=i+1}^{n-1}\left(W_{ij}\frac{s_i}{s_j}+W_{ji}\frac{s_j}{s_i}-W_{ij}-W_{ji}\right)\Bigg]G(s_1,\dots,s_{n-1},t) \, . \label{eq-G-fn}
\eea
With the initial condition
\be
P(Z_1,\dots,Z_{n-1},t=0)=\delta_{Z_1,0}\cdots\,\delta_{Z_{n-1},0} \qquad\mbox{and thus}\qquad G(s_1,\dots,s_{n-1},t=0)=1 \, ,
\ee
the solution is found to be
\be\label{G-fn-sol}
G(s_1,\dots,s_{n-1},t)=\exp\Bigg\{\Bigg[\sum_{i=1}^{n-1}\left(W_{i0}s_i+\frac{W_{0i}}{s_i}-W_{i0}-W_{0i}\right)+\sum_{i=1}^{n-1}\sum_{j=i+1}^{n-1}\left(W_{ij}\frac{s_i}{s_j}+W_{ji}\frac{s_j}{s_i}-W_{ij}-W_{ji}\right)\Bigg]t\Bigg\} .
\ee

If we denote by $N_{ij}$ the particle numbers transferred during the time interval $t$ between the reservoirs $R_i$ and $R_j$,  the number of particle transfers from reservoir $R_i$ can be expressed as
\be\label{Z(N)}
Z_i({\bf N})=\sum_{j=0,j\neq i}^{n-1}\left(N_{ij}-N_{ji}\right) \hspace{1cm} i=1,\dots,n-1 \, ,
\ee
where ${\bf N}=\{N_{ij}\}$.
The mean value of the number $N_{ij}$ is given by $\langle N_{ij}\rangle_t = W_{ij} t$ in terms of the corresponding rate $W_{ij}$.  These numbers $N_{ij}$ can be supposed to have Poisson distributions
\be\label{Poisson}
P(N_{ij},t) = {\rm e}^{-\langle N_{ij}\rangle_t} \, \frac{ \langle N_{ij}\rangle_t^{N_{ij}}}{N_{ij}!} \, .
\ee
Now, the probability distribution ruled by the master equation~(\ref{eq_general_master_equation}) can be written as
\be\label{multiPoisson}
P(Z_1,\dots,Z_{n-1},t) = \sum_{\bf N} \prod_i \delta_{Z_i,Z_i({\bf N})}  \, \prod_{i\ne j} P(N_{ij},t)  \, .
\ee
In this case, the generating function~(\ref{G-fn}) is indeed given by the solution~(\ref{G-fn-sol}) of Eq.~(\ref{eq-G-fn}), as can be verified by direct calculation.  Therefore, the stochastic process is here a combination of several independent Poisson processes.

Subsequently, we can obtain the cumulant generating function
\bea
Q(\lambda_1,\dots,\lambda_{n-1}) &=&\lim_{t\to\infty}-\frac{1}{t}\, \ln G\left({\rm e}^{-\lambda_1},\dots,{\rm e}^{-\lambda_{n-1}},t\right) \nonumber \\
&=&\sum_{i=1}^{n-1}\left[W_{i0}\left(1-{\rm e}^{-\lambda_i}\right)+W_{0i}\left(1-{\rm e}^{\lambda_i}\right)\right]+\sum_{i=1}^{n-1}\sum_{j=i+1}^{n-1}\left[W_{ij}\left(1-{\rm e}^{-\lambda_i+\lambda_j}\right)+W_{ji}\left(1-{\rm e}^{\lambda_i -\lambda_j}\right)\right] , \quad \label{eq_general_cumulant_generating_function}
\eea
which, under the condition of our hypothesis, can be regarded as a general form of cumulant generating function for any specific system in the sense that every possible transitions are allowed between the reservoirs.  However, the rates do not depend on internal states and systems that are more general in this regard will be considered below. In the following Section~\ref{Model2}, we show how the cumulant generating function~(\ref{eq_general_cumulant_generating_function}) can  also be obtained for a different model with the suitable identification of the rates $\{W_{ij}\}$.

With the above general cumulant generating function~(\ref{eq_general_cumulant_generating_function}), we have the symmetry relation
\be
Q(\lambda_1,\dots,\lambda_{n-1})=Q(A_1-\lambda_1,\dots,A_{n-1}-\lambda_{n-1}) \, , \label{eq_symmetry_relation}
\ee
with
\be\label{Aff-W}
A_i=\ln\frac{W_{i0}}{W_{0i}} \, ,
\ee
if the following conditions are satisfied
\be
A_i-A_j=\ln\frac{W_{ij}}{W_{ji}}\, . \label{eq_required_condition}
\ee

\subsection{The affinities and the central limit theorem}

In Eq.~(\ref{Aff-W}), $A_i$ is identified to be the affinity between the reservoirs $R_i$ and $R_0$ (this latter being considered as the reference reservoir).  In general, the affinity between the reservoirs $R_i$ and $R_j$ could be defined as
\be
A_{ij}=\ln\frac{W_{ij}}{W_{ji}}\, \, . \label{eq_affinity}
\ee
In order to obtain the affinities between any two reservoirs, we must first determine the value of the transition rates $\{W_{ij}\}$. There are as many such transition rates as
\be
S=n^2-n \, .
\ee
Our task is thus to find $S$ conditions, from which the value of $\{W_{ij}\}$ can be determined.

On the one hand, the conditions~(\ref{eq_required_condition}) give the following relations between the affinities,
\be
A_{ij}+A_{jk}=A_{ik}  \, , \label{eq_affinity_relation}
\ee
which equivalently leads to
\be
W_{ij}W_{jk}W_{ki}-W_{ji}W_{kj}W_{ik}=0 \, .
\ee
For a system with $n$ reservoirs, we can write down
\be\label{S1}
S_1=\frac{1}{2}\left(n^2-3n+2\right)
\ee
independent such affinity relations~(\ref{eq_affinity_relation}), which constitute a first group of conditions. We mathematically address how to find $S_1$ such conditions in Appendix \ref{app_affinity_relations}. 

On the other hand, according to the central limit theorem, the distribution of the particle exchanges ${\bf Z}$ during time interval $t$ is given by the Gaussian form~(\ref{eq_Gaussian_distribution}).  Taking the limit where $|Z_i|\gg 1$ (for $i=1,\dots,n-1$) in the master equation~(\ref{eq_general_master_equation}), we find that this Gaussian probability density ${\mathscr P}({\bf Z},t)$ should obey the following generalized Langevin equation
\be
\partial_t{\mathscr P}\simeq -\sum_{i=1}^{n-1}\sum_{j=0}^{n-1}\left(W_{ij}-W_{ji}\right)\partial_{Z_i}{\mathscr P} +\frac{1}{2}\sum_{i=1}^{n-1}\sum_{j=0,j\neq i}^{n-1}\left(W_{ij}+W_{ji}\right)\partial_{Z_i}^2{\mathscr P}  -\frac{1}{2} \sum_{i=1}^{n-1}\sum_{j=1,j\neq i}^{n-1}\left(W_{ij}+W_{ji}\right)\partial_{Z_i}\partial_{Z_j}{\mathscr P} \, .
\ee
Accordingly, we should have that
\bea
&& J_i=\sum_{j=0,j\neq i}^{n-1}\left(W_{ij}-W_{ji}\right) \hspace{1.5cm}\mbox{for}\quad  i=1,\dots,n-1 \, , \label{J_i}\\
&& D_{ii}=\frac{1}{2}\sum_{j=0,j\neq i}^{n-1}\left(W_{ij}+W_{ji}\right)\hspace{1cm}\mbox{for}\quad i=1,\dots,n-1 \, , \label{D_ii}\\
&& D_{ij}=-\frac{1}{2}\left(W_{ij}+W_{ji}\right)=D_{ji}\hspace{0.8cm}\mbox{for}\quad i,j=1,\dots,n-1; \ i\neq j \, .\label{D_ij}
\eea
These relations constitute a second group of 
\be
S_2=\frac{1}{2}\left(n^2+n-2\right)
\ee
independent conditions. Surprisingly, it can be easily verified that
\be
S_1+S_2=S \, .
\ee
Hence, all the independent conditions that are needed to determine the transition rates $\{W_{ij}\}$ are obtained. When there are more than two reservoirs in contact with the system, these $S$ conditions lead to so many nonlinear equations which can be solved numerically with the Newton-Raphson method to find all the rates $\{W_{ij}\}$. Therefore, the affinities between any two reservoirs can be evaluated by Eq.~(\ref{eq_affinity}). The method introduced here above has already been used to indirectly test the fluctuation theorem respectively in a two-reservoir system \cite{GG18} and a three-reservoir system \cite{GG19}. When evaluating the affinities, we now compare the computational/experimental expenses between this indirect method and the direct method which is based on the fluctuation relation, supposing that both methods are operationally feasible. Clearly, the indirect method developed here above is much cheaper since it is using a finite number of quantities, although the direct method is instead using a whole probability distribution.  The conclusion is that, for the model with constant rate, the affinities can be recovered from the knowledge of the first and second cumulants obtained in the long-time limit according to the central limit theorem.  The reason is that the system is compatible with detailed balance at equilibrium and the rates can thus be completely determined from the mean values of the currents and the diffusivities.

\vskip 0.5 cm

\subsection{Proof of consistency for systems near equilibrium}

Now, we prove that the method developed in previous subsection is exactly valid for any system near equilibrium. If the system is in equilibrium, then the affinities between any two reservoirs are equal to zero, and the currents are vanishing. We now apply a small perturbation to this system through small changes in the affinities. Under such circumstances, the currents can be expressed as linear responses of affinities, i.e.,
\be
\delta J_i=\sum_{j=1}^{n-1}L_{ij}\, \delta {\cal A}_j \, ,
\ee
where $L_{ij}$ are the linear response coefficients and ${\cal A}_j$ denote actual affinities. Here, we have reasonably omitted the nonlinear terms due to contributions that are negligible close enough to equilibrium. The Green-Kubo formulae state that $L_{ij}=D_{ij}$, where $\{D_{ij}\}$ are the diffusion coefficients for the system in equilibrium. These formulae are implied from the symmetry relation~(\ref{eq_symmetry_relation}). We thus have
\be
\delta J_i=\sum_{j=1}^{n-1}D_{ij}\, \delta {\cal A}_j \, . \label{eq_linear_response_a}
\ee
Let us now turn to the coarse-grained formalism given in previous subsection. The equilibrium condition implies that
\be
D_{ii} = \sum_{j=0, j\ne i}^{n-1} W_{ij} \qquad\mbox{and}\qquad D_{ij}=-W_{ij}=-W_{ji} \, , \label{eq_equilibrium_condition}
\ee
so that the variation of $A_{ij}\equiv\ln(W_{ij}/W_{ji})$ is as follows,
\be
\delta A_{ij}=\delta\left(\ln\frac{W_{ij}}{W_{ji}}\right)=\frac{\delta W_{ij}}{W_{ij}}-\frac{\delta W_{ji}}{W_{ji}}=\frac{1}{W_{ij}}\left(\delta W_{ij}-\delta W_{ji}\right) . \label{eq_variation_A}
\ee
With Eqs.~(\ref{eq_equilibrium_condition}) and~(\ref{eq_variation_A}), $\delta J_i$ can also be expressed as
\be
\delta J_i =\sum_{j=0}^{n-1}\left(\delta W_{ij}-\delta W_{ji}\right)=\sum_{j=0}^{n-1} W_{ij}\, \frac{\left(\delta W_{ij}-\delta W_{ji}\right)}{W_{ij}}=\sum_{j=0}^{n-1} W_{ij}\, \delta A_{ij} \, .
\ee
Because of $\delta A_{ij}=\delta A_i-\delta A_j$, $\delta J_i$ can be further expressed as
\be
\delta J_i=\sum_{j=0}^{n-1}W_{ij}\left(\delta A_i-\delta A_j\right)=\sum_{j=0, j\ne i}^{n-1}W_{ij}\, \delta A_i-\sum_{j=0, j\ne i}^{n-1}W_{ij} \, \delta A_j\, .
\ee
Due to Eq.~(\ref{eq_equilibrium_condition}), we reach the final desired form of $\delta J_i$, i.e.,
\be
\delta J_i=D_{ii}\, \delta A_i +\sum_{j=1, j\neq i}^{n-1}D_{ij}\, \delta A_j=\sum_{j=1}^{n-1}D_{ij}\, \delta A_j \, . \label{eq_linear_response_b}
\ee
Comparing Eq.~(\ref{eq_linear_response_b}) with Eq.~(\ref{eq_linear_response_a}), we immediately conclude that
\be
\delta A_i=\delta {\cal A}_i \, ,
\ee
which means that the quantities $A_i$ evaluated as $\ln(W_{i0}/W_{0i})$ can be identified as the actual affinities for the system near equilibrium, which proves the statement.

\subsection{Finite-time fluctuation relation}

The fluctuation relation can be directly proved starting from the expression~(\ref{multiPoisson}) and using the change of summation variables, $N_{ij}=N_{ji}'$.  As a consequence, we have that Eq.~(\ref{Z(N)}) is equivalent to $Z_i=-Z_i({\bf N}')=\sum_{j(\ne i)} (N_{ji}'-N_{ij}')$ in Eq.~(\ref{Z(N)}).  Therefore, using Eqs.~(\ref{eq_required_condition}) and~(\ref{Poisson}), Eq.~(\ref{multiPoisson}) becomes
\be
P({\bf Z},t) =  \sum_{{\bf N}'} \prod_i \delta_{-Z_i,Z_i({\bf N}')}  \, {\rm e}^{-\sum_{i\ne j} W_{ij}t} \, \prod_{i\ne j}  \frac{(W_{ji}t)^{N_{ji}'}}{N_{ji}'!} \, {\rm e}^{\sum_{i\ne j} (A_i-A_j)N_{ji}'} \, .
\ee
Now, we have that
\be
\sum_{i\ne j} (A_i-A_j)N_{ji}' = \sum_{i} A_i \sum_{j(\ne i)} (N_{ji}'-N_{ij}') = \sum_i A_i Z_i = {\bf A}\cdot{\bf Z} \, ,
\ee
hence the fluctuation relation
\be\label{FT_model1}
\frac{P({\bf Z},t)}{P(-{\bf Z},t)}  =  {\rm e}^{{\bf A}\cdot{\bf Z}}
\ee
is exactly satisfied at every instant of time.  As a consequence, the fluctuation relation~(\ref{eq_fluctuation_relation}) is satisfied in the long-time limit, as well as the symmetry relation~(\ref{FR_Q}) of the cumulant generating function.  Nevertheless, this model with constant rates is thus satisfying the stronger finite-time fluctuation relation~(\ref{FT_model1}) holding at every time and, moreover, for the affinities~(\ref{Aff-dfn}) that are constant in time.

\subsection{Time-reversal symmetry relations for the response properties}
\label{resp_prop_model1}

The time-reversal symmetry relations of Subsection~\ref{Gen-resp_coeff} can be obtained for the linear and nonlinear response coefficients of the model with constant rates.  The mean currents~(\ref{J_i}) can be expressed in terms of the affinities given by Eqs.~(\ref{Aff-W}) and~(\ref{eq_required_condition}) according to
\be
J_i = W_{0i}\left( {\rm e}^{A_i}-1\right) + \sum_{k>i} W_{ki}\left( {\rm e}^{A_i-A_k}-1\right) - \sum_{k<i} W_{ik}\left({\rm e}^{A_k-A_i}-1\right) . \label{J_i-Aff}
\ee
The diffusivities~(\ref{D_ii}) and~(\ref{D_ij}) are similarly given by
\bea
&& D_{ii}=\frac{1}{2}\, W_{0i}\left( {\rm e}^{A_i}+1\right) + \frac{1}{2}\sum_{k>i} W_{ki}\left( {\rm e}^{A_i-A_k}+1\right) + \frac{1}{2}\sum_{k<i} W_{ik}\left({\rm e}^{A_k-A_i}+1\right) ,  \label{D_ii-Aff}\\
&& D_{ij}=-\frac{1}{2}\, W_{ji} \left({\rm e}^{A_i-A_j}+1\right) \qquad\mbox{for}\quad i < j \, .\label{D_ij-Aff}
\eea

The linear response coefficients~(\ref{L_ij}) are thus taking the following values,
\bea
&& L_{i,i} = W_{0i} + \sum_{k>i} W_{ki} + \sum_{k<i} W_{ik} \, , \\
&& L_{i,j} = L_{j,i} = -W_{ji} \qquad \mbox{for} \quad i<j \, , 
\eea
so that the fluctuation-dissipation relations~(\ref{eq_FD_relation}) and Onsager's reciprocal relations~(\ref{eq_Onsager_reciprocal_relations}) are satisfied in the linear regime close to equilibrium.

Beyond, the nonlinear response coefficients~(\ref{M_ijk}) have the following expressions,
\bea
&& M_{i,ii} = W_{0i} + \sum_{k>i} W_{ki} - \sum_{k<i} W_{ik} \, , \\
&& M_{i,ij} = M_{j,ii} = -M_{i,jj} = -M_{j,ji} = -W_{ji} \qquad \mbox{for} \quad i<j \, , \\
&& M_{i,jk} = 0  \qquad\qquad\qquad\qquad\qquad\qquad\qquad\qquad \mbox{for} \quad i\ne j \ne k \, ,
\eea
while the first responses~(\ref{R_ijk}) of the diffusivities are here given by
\bea
&& R_{ii,i} = \frac{1}{2}\, W_{0i} + \frac{1}{2}\, \sum_{k>i} W_{ki} - \frac{1}{2}\, \sum_{k<i} W_{ik} \, , \\
&& R_{ii,j} = R_{ij,i} = -R_{ij,j} = -R_{jj,i} = -\frac{1}{2}\,W_{ji} \qquad\ \mbox{for} \quad i<j \, , \\
&& R_{ij,k} = 0  \qquad\qquad\qquad\qquad\qquad\qquad\qquad\qquad \ \; \mbox{for} \quad i\ne j \ne k \, ,
\eea
so that the symmetry relations~(\ref{eq_nonlinear_response_relations}) are also satisfied, since $M_{i,ii}=2R_{ii,i}$ for $i=j=k$, $M_{i,jj}=2R_{ij,j}$ for $j=k$, $M_{i,ij}=R_{ii,j}+R_{ij,i}$ for $k=i$, and $0=0$ for $i\ne j\ne k$.
These results confirm for the model with constant rates that the nonlinear response coefficients of the mean currents can be expressed in terms of the first responses of the diffusivities, as a consequence of the multivariate fluctuation relation~(\ref{eq_fluctuation_relation}).  These results will be used in the next section devoted to a more complicated model.

\section{Model with linear rates}
\label{Model2}

In this section, we consider a further stochastic model of transistor where the transition rates linearly depend on an internal random variable $N$, representing the occupancy of the system by particles.

\subsection{Master equation}

We first use a simple Markovian stochastic model as an example to show how we evaluate affinities. Let us consider the system in contact with three particle reservoirs, respectively with $\langle R_0\rangle$, $\langle R_1\rangle$, and $\langle R_2\rangle$ mean numbers of particles. The whole system can be schematically depicted by the following kinetic network
\begin{equation}
\begin{array}{ccccc}
R_1 & \autorightleftharpoons{$\scriptstyle k_{+1}$}{$\scriptstyle k_{-1}$} & N & \autorightleftharpoons{$\scriptstyle k_{-0}$}{$\scriptstyle k_{+0}$} & R_0 \\
& &  {\scriptstyle k_{+2}}\upharpoonleft\downharpoonright{\scriptstyle k_{-2}} & &  \\ & & & & \\
& &   R_2  & &
\end{array}
\label{eq_network}
\end{equation}
with the transition rates given by
\bea
&& W_{+0}(N)=k_{+0}\langle R_0\rangle\, ,\hspace{1cm} W_{-0}(N)=k_{-0}N \, , \\
&& W_{+1}(N)=k_{+1}\langle R_1\rangle\, ,\hspace{1cm} W_{-1}(N)=k_{-1}N \, , \\
&& W_{+2}(N)=k_{+2}\langle R_2\rangle\, ,\hspace{1cm} W_{-2}(N)=k_{-2}N \, .
\eea
The charging rates $\{W_{+i}\}$ are independent of the internal state $N$ of the system because they are all determined by the particle concentration of corresponding reservoir.  In contrast, the discharging rates $\{W_{-i}\}$ have a linear dependence on the system state $N$.

The probability distribution of the internal state $N$ of the system is governed by the master equation
\be
\frac{\rm d}{{\rm d}t}P(N,t) =\sum_{i=0}^2\left[\left({\rm e}^{-\partial_N}-1\right) W_{+i}(N) + \left({\rm e}^{+\partial_N}-1\right) W_{-i}(N)\right] P(N,t) , \label{eq_master_equation_10}
\ee
which has the following explicit form,
\be
\frac{\rm d}{{\rm d}t}P(N,t) =\sum_{i=0}^2\Bigg[k_{+i}\langle R_i\rangle P(N-1,t)+ k_{-i}(N+1)P(N+1,t)-\bigg(k_{+i}\langle R_i\rangle+k_{-i}N\bigg)P(N,t)\Bigg] . \label{eq_master_equation_1}
\ee

The evolution equation for the mean value $\langle N\rangle$ can be deduced from the master equation~(\ref{eq_master_equation_1}) to get
\be\label{kin_eq_1}
\frac{{\rm d}}{{\rm d}t}\langle N\rangle = \left(k_{+0}\langle R_0\rangle+k_{+1}\langle R_1\rangle+k_{+2}\langle R_2\rangle\right) - \left(k_{-0}+k_{-1}+k_{-2}\right) \langle N \rangle \, .
\ee
Consequently, the mean value in the steady state is given by
\be
\langle N \rangle_{\rm st} = \frac{k_{+0}\langle R_0\rangle+k_{+1}\langle R_1\rangle+k_{+2}\langle R_2\rangle}{k_{-0}+k_{-1}+k_{-2}} \, .
\ee
Since the rates $\{W_{-i}\}$ are linear functions of the system state $N$, the kinetic equation~(\ref{kin_eq_1}) is linear. The stationary solution of the master equation~(\ref{eq_master_equation_1}) is thus given by the following Poisson probability distribution,
\be\label{Poisson_st_st}
P_{\rm st}(N) = {\rm e}^{-\langle N\rangle_{\rm st}} \, \frac{ \langle N\rangle_{\rm st}^{N}}{N!} \, .
\ee

\subsection{Graph analysis and affinities}

\begin{figure}[h]
\begin{minipage}[t]{0.5\hsize}
\resizebox{1.0\hsize}{!}{\includegraphics{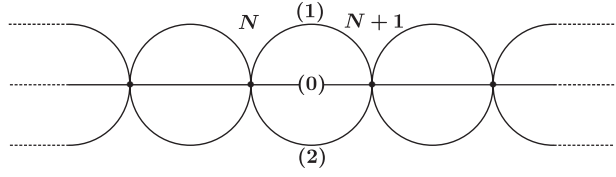}}
\end{minipage}
\caption{Graph associated with the Markovian stochastic model (\ref{eq_network}).}
\label{fig_graph}
\end{figure}

The graph associated with the above kinetic network (\ref{eq_network}) is depicted in Fig. \ref{fig_graph}. Each state of the system corresponds to a vertex and all the transitions between states are represented by edges. According to Schnakenberg's graph analysis \cite{S78}, the affinities can be identified by taking the logarithm of ratio of the products of transition rates along a cyclic path to those along the reversed path. Taking the event of a particle transfer from the reservoir $R_1$ to the reservoir $R_0$ as an example, the cyclic path and its reversed path are given by
\bea
&& c\equiv (N)\xrightarrow{W_{+1}}(N+1)\xrightarrow{W_{-0}}(N) \label{eq_path} \, , \\
&& c^{\rm r}\equiv (N)\xrightarrow{W_{+0}}(N+1)\xrightarrow{W_{-1}}(N) \label{eq_reserved_path} \, ,
\eea
which yield the affinity between the reservoirs $R_1$ and $R_0$ to be
\be
A_{10}=\ln\frac{W_{+1}(N)\, W_{-0}(N+1)}{W_{+0}(N)\, W_{-1}(N+1)}=\ln\frac{k_{+1}\langle R_1\rangle k_{-0}}{k_{+0}\langle R_0\rangle k_{-1}}\, .
\label{eq_theoretical_affinity_1}
\ee
Similarly, we have
\be
A_{20}=\ln\frac{k_{+2}\langle R_2\rangle k_{-0}}{k_{+0}\langle R_0\rangle k_{-2}}\, ,
\label{eq_theoretical_affinity_2}
\ee
which is the affinity between the reservoirs $R_2$ and $R_0$.  In the following, the affinity given by Schnakenberg's graph analysis will be called the theoretical affinity.

\subsection{Counting statistics}

For the Markovian process~(\ref{eq_network}), let us introduce the probability distribution $P(N,Z_1,Z_2,t)$ to have $N$ particles in the system and the signed cumulated fluxes $Z_1$ and $Z_2$, respectively from the reservoirs $R_1$ and $R_2$ towards the system, at time $t$.  This probability distribution evolves in time according the following master equation
\bea
\frac{\rm d}{{\rm d}t}P(N,Z_1,Z_2,t) &=&  \Big[\left({\rm e}^{-\partial_N}-1\right) k_{+0}\langle R_0\rangle +\left({\rm e}^{+\partial_N}-1\right)k_{-0}N  \nonumber \\
&& +\left({\rm e}^{-\partial_N}{\rm e}^{-\partial_{Z_1}}-1\right) k_{+1}\langle R_1\rangle +\left({\rm e}^{+\partial_N}{\rm e}^{+\partial_{Z_1}}-1\right) k_{-1}N
 \nonumber \\
&& +\left({\rm e}^{-\partial_N}{\rm e}^{-\partial_{Z_2}}-1\right) k_{+2}\langle R_2\rangle +\left({\rm e}^{+\partial_N}{\rm e}^{+\partial_{Z_2}}-1\right) k_{-2}N \Big] P(N,Z_1,Z_2,t) \, .
\eea
Let us define the finite-time generating function of the signed cumulated fluxes \cite{G04}, reading
\be
G(s,s_1,s_2,t)\equiv\sum_{N=0}^{\infty}\sum_{Z_1,Z_2=-\infty}^{+\infty}s^N s_1^{Z_1}s_2^{Z_2}P(N,Z_1,Z_2,t) \, ,
\ee
and set the initial condition according to the steady state, i.e.,
\be
P(N,Z_1,Z_2,t=0)={\rm e}^{-\langle N\rangle_{\rm st}}\frac{\langle N\rangle_{\rm st}^N}{N!}\, \delta_{Z_1,0}\, \delta_{Z_2,0} \, ,
\ee
so that
\be
G(s,s_1,s_2,t=0)={\rm e}^{\langle N\rangle_{\rm st}(s-1)} \, .
\label{IC_G}
\ee
Then, from the above master equation, we get the evolution equation for the generating function as
\bea\label{Eq-G-model2}
\partial_t  G(s,s_1,s_2,t) &=& \left[k_{+0}\langle R_0\rangle(s-1)+k_{+1}\langle R_1\rangle(ss_1-1)+k_{+2}\langle R_2\rangle(ss_2-1)\right]G(s,s_1,s_2,t) \nonumber \\
&& +  \left[k_{-0}(1-s)+k_{-1}\left(\frac{1}{s_1}-s\right)+k_{-2}\left(\frac{1}{s_2}-s\right)\right]\partial_s G(s,s_1,s_2,t) 
\, ,
\eea
which can be written in the following form,
\be
\partial_t  G = (As-B)\, G + (C-Ds)\, \partial_s  G
\ee
with
\bea
&& A=k_{+0}\langle R_0\rangle+k_{+1}\langle R_1\rangle s_1+k_{+2}\langle R_2\rangle s_2 \, , \label{coeff-A}\\
&& B=k_{+0}\langle R_0\rangle+k_{+1}\langle R_1\rangle+k_{+2}\langle R_2\rangle \, , \label{coeff-B}\\
&& C=k_{-0}+\frac{k_{-1}}{s_1}+\frac{k_{-2}}{s_2} \, , \label{coeff-C}\\
&& D=k_{-0}+k_{-1}+k_{-2} \, . \label{coeff-D}
\eea
We note that $B=A(s_1=s_2=1)$ and $D=C(s_1=s_2=1)$.
The first-order partial differential equation~(\ref{Eq-G-model2}) with the given initial condition can be solved using the method of characteristics \cite{G04}.  Accordingly, the problem reduces to solving the two ordinary differential equations:
\bea
&& \frac{{\rm d}G}{{\rm d}t} = (As-B) \, G \, , \\
&& \frac{{\rm d}s}{{\rm d}t} = -C + Ds \, ,
\eea
from the initial conditions $G_0$ and $s_0$ that are coupled together by Eq.~(\ref{IC_G}), which reads $G_0=\exp[\langle N\rangle_{\rm st}(s_0-1)]$ with $\langle N\rangle_{\rm st}=B/D$.  Integrating these ordinary differential equations and eliminating $G_0$ and $s_0$, we obtain the solution
\be\label{G_model2}
G(s,s_1,s_2,t) = \exp\left\{ t \, \frac{AC-BD}{D}-\frac{B}{D}-\frac{(A-B)C}{D^2}\left(1-{\rm e}^{-Dt}\right) + \left[\frac{B}{D}\, {\rm e}^{-Dt}+\frac{A}{D}\left(1-{\rm e}^{-Dt}\right)\right] s \right\} .
\ee

The cumulant generating function
\be
Q(\lambda_1,\lambda_2)  =\lim_{t\to\infty}-\frac{1}{t}\, \ln G\left(1,{\rm e}^{-\lambda_1},{\rm e}^{-\lambda_2},t\right)  =\frac{{B}{D}-{A}{C}}{D}
\label{eq_generating_function}
\ee
is therefore obtained with
\bea
&& {A}=k_{+0}\langle R_0\rangle+k_{+1}\langle R_1\rangle\, {\rm e}^{-\lambda_1}+k_{+2}\langle R_2\rangle\, {\rm e}^{-\lambda_2} \, , \\
&& {B}=k_{+0}\langle R_0\rangle+k_{+1}\langle R_1\rangle+k_{+2}\langle R_2\rangle \, , \\
&& {C}=k_{-0}+k_{-1}\, {\rm e}^{\lambda_1}+k_{-2}\, {\rm e}^{\lambda_2} \, , \\
&& {D}=k_{-0}+k_{-1}+k_{-2} \, .
\eea
If we make the following identification
\bea
&& W_{10} \equiv \frac{k_{+1}\langle R_1\rangle k_{-0}}{k_{-0}+k_{-1}+k_{-2}}\, ,\hspace{1cm} W_{01} \equiv \frac{k_{+0}\langle R_0\rangle k_{-1}}{k_{-0}+k_{-1}+k_{-2}}\, ,\label{eq_rate_1} \\
&& W_{20} \equiv \frac{k_{+2}\langle R_2\rangle k_{-0}}{k_{-0}+k_{-1}+k_{-2}}\, ,\hspace{1cm} W_{02} \equiv \frac{k_{+0}\langle R_0\rangle k_{-2}}{k_{-0}+k_{-1}+k_{-2}}\, ,\label{eq_rate_2} \\
&& W_{12} \equiv \frac{k_{+1}\langle R_1\rangle k_{-2}}{k_{-0}+k_{-1}+k_{-2}}\, ,\hspace{1cm} W_{21} \equiv \frac{k_{+2}\langle R_2\rangle k_{-1}}{k_{-0}+k_{-1}+k_{-2}}\, ,\label{eq_rate_3} 
\eea
then the cumulant generating function Eq.~(\ref{eq_generating_function}) becomes
\bea
Q(\lambda_1,\lambda_2)&=& W_{10}\left(1-{\rm e}^{-\lambda_1}\right)+W_{20}\left(1-{\rm e}^{-\lambda_2}\right)+W_{12}\left(1- {\rm e}^{-\lambda_1+\lambda_2}\right)\nonumber\\
&& +W_{01}\left(1-{\rm e}^{\lambda_1}\right)+W_{02}\left(1-{\rm e}^{\lambda_2}\right) +W_{21}\left(1-{\rm e}^{\lambda_1-\lambda_2}\right) ,  \label{Q_model2}
\eea
which has the same form as Eq.~(\ref{eq_general_cumulant_generating_function}) with $n=3$. Accordingly, the quantities given by Eqs.~(\ref{eq_rate_1}), (\ref{eq_rate_2}), and~(\ref{eq_rate_3}) are the equivalent transition rates between the different reservoirs.  Therefore, the symmetry relation~(\ref{eq_symmetry_relation}) is here also satisfied,
\be
Q(\lambda_1,\lambda_{2})=Q(A_{10}-\lambda_1,A_{20}-\lambda_{2}) \, , \label{QFR_model2}
\ee
with the following affinities,
\bea
&& A_{10}=\ln\frac{W_{10}}{W_{01}}=\ln\frac{k_{+1}\langle R_1\rangle k_{-0}}{k_{+0}\langle R_0\rangle k_{-1}}\, \, , \\
&& A_{20}=\ln\frac{W_{20}}{W_{02}}=\ln\frac{k_{+2}\langle R_2\rangle k_{-0}}{k_{+0}\langle R_0\rangle k_{-2}}\, \, ,
\eea
which are equivalent to those of Eqs.~(\ref{eq_theoretical_affinity_1}) and~(\ref{eq_theoretical_affinity_2}) given by Schnakenberg's graph analysis. Furthermore, the  fact that the affinities can be expressed in terms of the rates defined by Eqs.~(\ref{eq_rate_1})-(\ref{eq_rate_3}) in a similar way as for the model with constant rates supports the hypothesis according to which the affinities can be determined from the knowledge of the mean currents and diffusivities obtained by statistics over an arbitrarily long time interval.

The Markovian stochastic process ruled by Eq.~(\ref{eq_master_equation_1}) can be exactly simulated using Gillespie's algorithm \cite{G76}. However, it is computationally expensive since the particles transit one by one. A faster simulation method is provided by the corresponding Langevin stochastic process whose procedure is presented in Appendix~\ref{app_Langevin_stochastic_process}. In the simulations, we perform the counting statistics of particle transfers $Z_1$ and $Z_2$, respectively  from the reservoirs $R_1$ and $R_2$ to the system during a long enough time interval $t$. The joint distribution of $Z_1$ and $Z_2$ is well approximated by the Gaussian distribution (\ref{eq_Gaussian_distribution}), from which we obtain the numerical value of $J_1$, $J_2$, $D_{11}$, $D_{22}$, and $D_{12}$. Gathering all the independent conditions, we have the nonlinear equations
\bea
&& W_{10}-W_{01}+W_{12}-W_{21}=J_1 \, , \label{W-J1}\\
&& W_{20}-W_{02}-W_{12}+W_{21}=J_2 \, ,  \label{W-J2}\\
&& W_{10}+W_{01}+W_{12}+W_{21}=2D_{11} \, ,  \label{W-D11}\\
&& W_{20}+W_{02}+W_{12}+W_{21}=2D_{22} \, , \label{W-D22}\\
&& W_{12}+W_{21}=-2D_{12} \, , \label{W-D12}\\
&& W_{01}W_{12}W_{20}=W_{02}W_{21}W_{10} \, , \label{WWW}
\eea
which can be numerically solved using the Newton-Raphson method. After finding the roots of this nonlinear equations, we can readily obtain the numerical values of the affinities through
\bea
&& A_{10}=\ln\frac{W_{10}}{W_{01}}\, , \label{A10}\\
&& A_{20}=\ln\frac{W_{20}}{W_{02}}\, . \label{A20}
\eea
From now on, the affinity evaluated through the counting-statistical method will be called the numerical affinity. In Table~\ref{tab_a_1}, we present the comparison of numerical and theoretical affinities for the system under different conditions. The agreement between the values confirms the control of the counting statistics by the first and second cumulants for the transistor model with linear rates.

\begin{table}[h]
\caption{Numerical results for the linear model of transistor. The numerical and theoretical affinities are compared  for different conditions in the reservoirs. The mean values of the particle numbers in the reservoirs are denoted $\langle R_0\rangle$, $\langle R_1\rangle$, and $\langle R_2\rangle$. We set $k_{\pm 0}=k_{\pm 1}=k_{\pm 2}=1$ in numerical simulations and the numerical affinities are computed over the time interval $t=100$ with $2\times 10^5$ data.}
\vskip 0.3 cm
\begin{tabular}{rrrrrrr}
\hline
\hline
$\qquad\langle R_0\rangle$ & $\qquad\langle R_1\rangle$ & $\qquad\langle R_2\rangle$ & $\qquad A_{10}^{\rm (th)}$ & $\qquad A_{10}^{\rm (num)}\qquad$ & $\qquad A_{20}^{\rm (th)}$ & $\qquad A_{20}^{\rm (num)}\qquad$ \bigstrut \\ \hline
$1000$ & $3000$ & $2000$ & $1.099$ & $1.089 \pm 0.008$ & $0.693$ & $0.684  \pm 0.007$ \bigstrut \\ 
$3000$ & $1000$ & $2000$ & $-1.099$ & $-1.096 \pm 0.005$ & $-0.405$ & $-0.403 \pm 0.003$ \bigstrut \\ 
$2000$ & $10000$ & $1000$ & $1.609$ & $1.595 \pm 0.012$ & $-0.693$ & $-0.706 \pm  0.011$ \bigstrut \\ 
$1000$ & $10000$ & $1000$ & $2.303$ & $2.270 \pm 0.023$ & $0.000$ & $-0.030 \pm 0.020$ \bigstrut \\ 
$10000$ & $2000$ & $1000$ & $-1.609$ & $-1.608 \pm 0.009$ & $-2.303$ & $-2.304  \pm 0.014$ \bigstrut \\ 
$15000$ & $2000$ & $1000$ & $-2.015$ & $-2.015 \pm  0.013$ & $-2.708$ & $-2.713 \pm 0.019$ \bigstrut \\ 
$3000$ & $20000$ & $1000$ & $1.897$ & $1.876 \pm 0.017$ & $-1.099$ & $-1.120 \pm 0.019$ \bigstrut \\ 
$1000$ & $1000$ & $15000$ & $0.000$ & $-0.041 \pm 0.031$ & $2.708$ & $2.651 \pm 0.035$ \bigstrut \\ 
$10000$ & $1000$ & $20000$ & $-2.303$ & $-2.301 \pm 0.024$ & $0.693$ & $0.689 \pm 0.005$ \bigstrut \\ 
\hline
\hline
\end{tabular}
\label{tab_a_1}
\end{table}

\subsection{Finite-time fluctuation relation}

We observe that the generating function~(\ref{G_model2}) with $s=1$ for the model with linear rates has the same structure as the function~(\ref{G-fn-sol}) for the model with constant rates.  This observation suggests that the stochastic exchange process is here also a combination of Poisson processes of the type~(\ref{Poisson}).  Indeed, since $Z_1=N_{10}-N_{01}+N_{12}-N_{21}$ and $Z_2=N_{20}-N_{02}-N_{12}+N_{21}$, we may write
\be
G(s=1,s_1,s_2,t) = \sum_{\{N_{ij}\}} s_1^{N_{10}-N_{01}+N_{12}-N_{21}} s_2^{N_{20}-N_{02}-N_{12}+N_{21}} \prod_{i\ne j} {\rm e}^{-\nu_{ij}} \frac{\nu_{ij}^{N_{ij}}}{N_{ij}!}
\ee
to obtain
\be
G(s=1,s_1,s_2,t) = \exp \left[\nu_{10}\left(s_1-1\right)+\nu_{01}\left(\frac{1}{s_1}-1\right)+\nu_{20}\left(s_2-1\right)+\nu_{02}\left(\frac{1}{s_2}-1\right)+ \nu_{12}\left(\frac{s_1}{s_2}-1\right)+\nu_{21}\left(\frac{s_2}{s_1}-1\right) \right] .
\ee
Comparing to the expression~(\ref{G_model2}) with $s=1$ and the coefficients~(\ref{coeff-A})-(\ref{coeff-D}),
we can identify the parameters $\{\nu_{ij}\}$ of the Poisson distributions as
\bea
&& \nu_{10} \equiv \langle N_{10}\rangle_t = W_{10}\, t + (W_{11}+W_{12}) \, \frac{f(t)}{D} \, , \label{nu_10}\\
&& \nu_{01} \equiv \langle N_{01}\rangle_t = W_{01}\, t + (W_{11}+W_{21}) \, \frac{f(t)}{D} \, , \label{nu_01}\\
&& \nu_{20} \equiv \langle N_{20}\rangle_t = W_{20}\, t + (W_{21}+W_{22}) \, \frac{f(t)}{D} \, , \label{nu_20}\\
&& \nu_{02} \equiv \langle N_{02}\rangle_t = W_{02}\, t + (W_{12}+W_{22}) \, \frac{f(t)}{D} \, , \label{nu_02}\\
&& \nu_{12} \equiv \langle N_{12}\rangle_t = W_{12}\left[ t -\frac{f(t)}{D}\right] , \label{nu_12}\\
&& \nu_{21} \equiv \langle N_{21}\rangle_t = W_{21}\left[ t -\frac{f(t)}{D}\right] , \label{nu_21}
\eea
in terms of the rates~(\ref{eq_rate_1})-(\ref{eq_rate_3}), the further quantities
\be\label{W_ii}
W_{ii}\equiv \frac{k_{+i}\langle R_i\rangle k_{-i}}{k_{-0}+k_{-1}+k_{-2}} \qquad\mbox{for}\quad i=1,2 \, ,
\ee
and the function
\be
f(t) = 1 - \exp(-Dt) \, . \label{fn-f}
\ee

Now, the definitions~(\ref{eq_rate_1})-(\ref{eq_rate_3}) and~(\ref{W_ii}) for the rates imply not only the relation~(\ref{WWW}), but also the similar relation
\be
\nu_{01}(t)\, \nu_{12}(t)\, \nu_{20}(t) = \nu_{02}(t)\, \nu_{21}(t) \, \nu_{10}(t)
\ee
between the time-dependent parameters~(\ref{nu_10})-(\ref{nu_21}) of the Poisson distributions.  If we introduce the time-dependent affinities as
\bea
&& \tilde A_{10}(t) \equiv \ln\frac{\nu_{10}(t)}{\nu_{01}(t)}\, , \label{A10t} \\
&& \tilde A_{20}(t) \equiv \ln\frac{\nu_{20}(t)}{\nu_{02}(t)}\, , \label{A20t} \\
&& \tilde A_{12}(t) \equiv \ln\frac{\nu_{12}(t)}{\nu_{21}(t)}\, , \label{A12t}
\eea
in terms of the parameters~(\ref{nu_10})-(\ref{nu_21}), we thus have the property that
\be
\Delta\tilde A(t) \equiv \tilde A_{12}(t)-\tilde A_{10}(t)+\tilde A_{20}(t) = 0 \, .
\label{DAt}
\ee

The probability distribution of the particle numbers exchanged between the three reservoirs,
\bea
&&\Delta N_{10} \equiv N_{10}-N_{01} \, , \\
&&\Delta N_{20} \equiv N_{20}-N_{02} \, , \\
&&\Delta N_{12} \equiv N_{12}-N_{21} \, , 
\eea
is given by
\be
P(\Delta N_{10},\Delta N_{20},\Delta N_{12},t) = \sum_{\{N_{ij}\}} \delta_{\Delta N_{10},N_{10}-N_{01}} \, \delta_{\Delta N_{20},N_{20}-N_{02}} \, \delta_{\Delta N_{12},N_{12}-N_{21}} \,  \prod_{i\ne j} {\rm e}^{-\nu_{ij}} \frac{\nu_{ij}^{N_{ij}}}{N_{ij}!} \, .
\ee
Comparing with the probability distribution for the opposite fluctuations of the particle numbers, we deduce the following finite-time trivariate symmetry relation
\be
\frac{P(\Delta N_{10},\Delta N_{20},\Delta N_{12},t)}{P(-\Delta N_{10},-\Delta N_{20},-\Delta N_{12},t)} = \exp\left[\tilde A_{10}(t) \, \Delta N_{10}+\tilde A_{20}(t) \, \Delta N_{20}+\tilde A_{12}(t) \, \Delta N_{12} \right],
\label{triple_FR}
\ee
which holds at every time in terms of the time-dependent affinities~(\ref{A10t})-(\ref{A12t}).  Since the numbers of particles exchanged between the reservoirs $i=1,2$ and the reference reservoir $i=0$ are given by $Z_1=\Delta N_{10}+\Delta N_{12}$ and $Z_2=\Delta N_{20}-\Delta N_{12}$, the finite-time trivariate fluctuation relation~(\ref{triple_FR}) can be written equivalently in the following form,
\be
\frac{P(Z_1,Z_2,\Delta N_{12},t)}{P(-Z_1,-Z_2,-\Delta N_{12},t)} = \exp\left[\tilde A_{10}(t)\, Z_1+\tilde A_{20}(t) \,Z_2+\Delta\tilde A(t) \, \Delta N_{12} \right] 
\label{triple_FR_bis}
\ee
in terms of the quantity~(\ref{DAt}), which is vanishing.  Consequently, the right-hand side of Eq.~(\ref{triple_FR_bis}) no longer depends on $\Delta N_{12}$.  Therefore, multiplying Eq.~(\ref{triple_FR_bis}) by $P(-Z_1,-Z_2,-\Delta N_{12},t)$ and summing over $\Delta N_{12}$ to form the marginal probability distribution $P(Z_1,Z_2,t)\equiv \sum_{\Delta N_{12}=-\infty}^{+\infty} P(Z_1,Z_2,\Delta N_{12},t)$, we obtain the bivariate fluctuation relation
\be\label{biFR_model2}
\frac{P(Z_1,Z_2,t)}{P(-Z_1,-Z_2,t)} = \exp\left[ \tilde A_{10}(t) \, Z_1+ \tilde A_{20}(t) \, Z_2 \right] ,
\ee
here also holding at every time with the finite-time affinities defined by Eqs.~(\ref{A10t}) and~(\ref{A20t}).
In the long-time limit, these affinities behave as $\tilde A_{i0}(t)=A_{i0}+O[(Dt)^{-1}]$, where $A_{i0}$ are the affinities~(\ref{A10}) and~(\ref{A20}) (for $i=1,2$).  Therefore, the asymptotic fluctuation relation
\be\label{biFR_model2_infty}
\frac{P(Z_1,Z_2,t)}{P(-Z_1,-Z_2,t)} \sim_{t\to\infty} \exp\left( A_{10} \, Z_1+ A_{20} \, Z_2 \right)
\ee
is recovered in the long-time limit, which is consistent with the validity of the symmetry relation~(\ref{QFR_model2}) satisfied by the cumulant generating function.  The analysis shows that the asymptotic symmetry~(\ref{biFR_model2_infty}) is slowly approached in time with corrections going as $t^{-1}$ and becoming negligible over time scales $t \gg D^{-1}=(k_{-0}+k_{-1}+k_{-2})^{-1}$.

\vskip 0.5 cm

\subsection{Time-reversal symmetry relations for the response properties}

As shown in Subsection~\ref{Gen-resp_coeff}, the symmetry relation~(\ref{QFR_model2}) implies the fluctuation-dissipation relations~(\ref{eq_FD_relation}) and the Onsager reciprocal relations~(\ref{eq_Onsager_reciprocal_relations}) in the linear regime close to equilibrium, as well as their generalizations such as Eq.~(\ref{eq_nonlinear_response_relations}) in the nonlinear regimes farther away from equilibrium.

Since the cumulant generating function~(\ref{Q_model2}) is here precisely of the same form as Eq.~(\ref{eq_general_cumulant_generating_function}) with $n=3$ for the model with constant rates, the response coefficients of the present model have the same expressions as those obtained in Subsection~\ref{resp_prop_model1} for the previous model.  We thus have that the linear response coefficients and the diffusivities satisfy Eqs.~(\ref{eq_FD_relation}) and~(\ref{eq_Onsager_reciprocal_relations}) because
\bea
&& L_{1,1} = W_{01}+W_{21} = D_{11}(0,0) \, , \\
&& L_{1,2}=L_{2,1}=-W_{21} = D_{12}(0,0)  = D_{21}(0,0)\, , \\
&& L_{2,2}=W_{02}+W_{21} = D_{22}(0,0) \, .
\eea
Moreover, the nonlinear response coefficients~(\ref{M_ijk}) are indeed related to the first responses~(\ref{R_ijk}) of the diffusivities since
\bea
&& M_{1,11} = W_{01}+W_{21} = 2 \, R_{11,1} \, , \qquad\  \, M_{2,11} = -W_{21} = 2 \, R_{21,1} \, , \\
&& M_{1,12} = -W_{21} = R_{11,2} + R_{12,1} \, , \qquad M_{2,12} = W_{21} = R_{21,2} + R_{22,1} \, , \\
&& M_{1,21} = -W_{21} = R_{12,1} + R_{11,2} \, , \qquad M_{2,21} = W_{21} = R_{22,1} + R_{21,2} \, , \\
&& M_{1,22} = W_{21} = 2\, R_{12,2} \, , \qquad\qquad\quad\ \, M_{2,22} = W_{02}-W_{21} = 2 \, R_{22,2} \, , 
\eea
as predicted by Eq.~(\ref{eq_nonlinear_response_relations}).

\section{Model with nonlinear rates}
\label{Model3}

In the present section, a further stochastic model of transistor is considered where the transition rates have a nonlinear dependence on the internal random occupancy $N$ of the system by particles.  The model of two mesoscopic tunnel junctions coupled in series described in Ref.~\cite{AWBMJ91} is extended to a model of transistor with three tunnel junctions~\cite{AG06JSM}.  This model can also describe single-electron transistors in the limit where some gate resistance becomes large enough \cite{AL86,K92}.

\subsection{Master equation}

\begin{figure}[h]
\begin{minipage}[t]{0.5\hsize}
\resizebox{1.0\hsize}{!}{\includegraphics{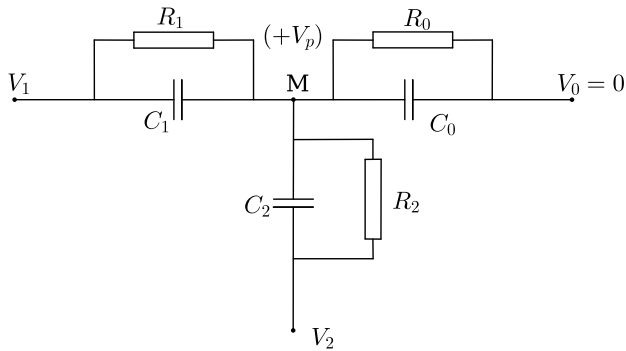}}
\end{minipage}
\caption{Schematic representation of three junctions, which are characterized by their respective resistances $\{R_0,R_1,R_2\}$ and capacitances $\{C_0,C_1,C_2\}$. Two leads are set to the voltages $V_1$ and $V_2$, while the other one is set fixed to $V_0=0$ as the reference. $V_p$ is included to account for any misalignment of Fermi level of the central region with respect to those of three leads when $V_1=V_2=V_0=0$.}
\label{fig_junction}
\end{figure}

We consider three mesoscopic tunnel junctions coupled with each other, as shown in Fig. \ref{fig_junction} \cite{AWBMJ91,AG06JSM}.  Every junction is characterized by its capacitance $C_i$ and its resistance $R_i$ ($i=0,1,2$).  The system is thus composed of a central region M, which is a conductive island such as a quantum dot, coupled through the mesoscopic tunnel junctions to three electron reservoirs at the voltages $V_i$.  In this mesoscopic system, the process of electron transport is stochastic.  The number $N$ of excess electrons in the conductive island is a random variable varying in time because of stochastic electron tunneling events across the junctions.  This number $N$ determines the state of conductive island in the semiclassical description of the system.

According to classical electrodynamics, the voltage of the central region is found to be
\be\label{V_M}
V_{\rm M}(N)=\frac{V_0C_0+V_1C_1+V_2C_2}{C_0+C_1+C_2}-\frac{Ne}{C_0+C_1+C_2}+V_p \, ,
\ee
where $V_p$ takes into account the possible misalignment of Fermi level in the central region with respect to the three leads when the junction system is in equilibrium, i.e., $V_1=V_2=V_0=0$.  This parameter can also be interpreted in terms of the background charge $q_0=-(C_0+C_1+C_2)V_p$ in the central region, or as a gate voltage.  We should note that the shift of the potential in discrete way by changing $N$ can offset the influence of $V_p$ on the junction system, so that only the value of
\be\label{V_p}
V_p \quad \text{modulo}\quad \frac{e}{C_0+C_1+C_2}
\ee
can serve as an indicator of the influence of $V_p$ on the junction system (beside the number of excess electrons in the central region).  As a consequence, the transport properties of the system manifest a periodic dependence on $V_p$, as will be shown below.

The time evolution of the probability $P(N,t)$ that the conductive island is occupied by $N$ excess electrons is ruled by the master equation
\be\label{master_eq_model3}
\frac{\rm d}{{\rm d}t}P(N,t)=\sum_{i=0,1,2}\sum_{\pm}\bigg[W_{i}^{(\pm)}(N\mp1)P(N\mp 1,t)-W_{i}^{(\pm)}(N)P(N,t)\bigg]  .
\ee
The transition rates can be obtained using Fermi's golden rule \cite{AWBMJ91} and they are given by
\be\label{rates_model3}
W_{i}^{(\pm)}(N)=\frac{1}{e^2R_{i}}\frac{\Delta E_{i}^{(\pm)}(N)}{\exp\left[\beta\Delta E_{i}^{(\pm)}(N)\right]-1} \, ,
\ee
where $\Delta E_{i}^{(\pm)}(N)$ denotes the energy jump of the system during electron tunneling charging (respectively discharging) the conductive island through the $i$th junction.  These jumps of energy can be calculated by considering the contributions from the change in the system electrostatic energy and from the work done by the external voltage in the transition \cite{AWBMJ91}.  They can be expressed as
\be\label{Delta_E}
\Delta E_{i}^{(\pm)}(N)=\mp e\left[V_{\rm M}(N)-V_{i}\right]+E_c
\ee
in terms of the capacitive charging energy $E_c=e^2/2C$, where $C=C_0+C_1+C_2$ is the total capacitance.
The phenomenon of Coulomb blockade manifests itself when this energy is larger than the thermal energy $k_{\rm B}T$.

The global electrostatic interaction is thus embodied in the dependence of the rates on the occupancy $N$ of the conductive island. The process has a nonlinear character because the transition rates are nonlinear functions of the excess electron number $N$ in the central region. For this reason, the probability distribution $P_{\rm st}(N)$ that is the stationary solution of the master equation~(\ref{master_eq_model3}) is no longer Poissonian in contrast to the situation in the model with linear rates.

\subsection{Counting statistics}
\label{Countstat_model3}

Here, the counting statistics can be obtained as for the previous models by considering the following extended master equation,
\bea
\frac{\rm d}{{\rm d}t}P(N,Z_1,Z_2,t) &=&  \Big[\left({\rm e}^{-\partial_N}-1\right) W_{0}^{(+)}(N) +\left({\rm e}^{+\partial_N}-1\right)W_{0}^{(-)}(N)  \nonumber \\
&& +\left({\rm e}^{-\partial_N}{\rm e}^{-\partial_{Z_1}}-1\right) W_{1}^{(+)}(N) +\left({\rm e}^{+\partial_N}{\rm e}^{+\partial_{Z_1}}-1\right) W_{1}^{(-)}(N)
 \nonumber \\
&& +\left({\rm e}^{-\partial_N}{\rm e}^{-\partial_{Z_2}}-1\right) W_{2}^{(+)}(N) +\left({\rm e}^{+\partial_N}{\rm e}^{+\partial_{Z_2}}-1\right) W_{2}^{(-)}(N) \Big] P(N,Z_1,Z_2,t) \, .
\eea
Now, the cumulant generating function $Q(\pmb{\lambda})$ with $\pmb{\lambda}=(\lambda_1,\lambda_2)$ can be obtained by solving the eigenvalue problem
\be\label{eigenvalue}
\hat{L}_{\pmb{\lambda}}\Psi_{\pmb{\lambda}}(N)=-Q(\pmb{\lambda})\Psi_{\pmb{\lambda}}(N) \, , \qquad
\hat{L}_{\pmb{\lambda}}^{\dagger}\tilde\Psi_{\pmb{\lambda}}(N)=-Q(\pmb{\lambda})\tilde\Psi_{\pmb{\lambda}}(N) \, ,
\ee
where the operator $\hat{L}_{\pmb{\lambda}}$ is defined through the equation
\bea
\frac{\rm d}{{\rm d}t}F(N,t) =\hat{L}_{\pmb{\lambda}} F(N,t) &=& \Big[\left({\rm e}^{-\partial_N}-1\right) W_{0}^{(+)}(N) +\left( {\rm e}^{+\partial_N}-1\right) W_{0}^{(-)}(N)   \nonumber \\
&& +\left( {\rm e}^{-\partial_N}{\rm e}^{-\lambda_1}-1\right) W_{1}^{(+)}(N) +\left( {\rm e}^{+\partial_N}{\rm e}^{+\lambda_1}-1\right) W_{1}^{(-)}(N) \nonumber \\
&& +\left( {\rm e}^{-\partial_N}{\rm e}^{-\lambda_2}-1\right) W_{2}^{(+)}(N) +\left( {\rm e}^{+\partial_N}{\rm e}^{+\lambda_2}-1\right) W_{2}^{(-)}(N)\Big] F(N,t) \, , \label{eq_modified_master_equation}
\eea
and $\hat{L}_{\pmb{\lambda}}^{\dagger}$ denotes its adjoint.  In Eq.~(\ref{eigenvalue}), $-Q({\pmb{\lambda}})$ is the leading eigenvalue, $\Psi_{\pmb{\lambda}}(N)$ the corresponding right-eigendistribution, and $\tilde\Psi_{\pmb{\lambda}}(N)$ the left-eigendistribution. The matrix elements of the operator $\hat{L}_{\pmb{\lambda}}$ are given by
\be
\hat{L}_{\pmb{\lambda}}(N,N')\equiv \Gamma_{\pmb{\lambda}}^{(+)}(N')\delta_{N-1,N'}+\Gamma_{\pmb{\lambda}}^{(-)}(N')\delta_{N+1,N'}-\left[\Gamma_{\bf 0}^{(+)}(N')+\Gamma_{\bf 0}^{(-)}(N')\right]\delta_{N,N'} \, ,
\ee
where
\be
\Gamma_{\pmb{\lambda}}^{(\pm)} (N)=W_{0}^{(\pm)}(N) + W_{1}^{(\pm)}(N)\, {\rm e}^{\mp\lambda_1} + W_{2}^{(\pm)}(N)\, {\rm e}^{\mp\lambda_2} \, .
\ee
We note that the cumulant generating function $Q(\pmb{\lambda})$ can thus be expressed as
\be
Q(\pmb{\lambda})=-\frac{\sum_{N,N'}\tilde\Psi_{\pmb{\lambda}}(N)\hat{L}_{\pmb{\lambda}}(N,N')\Psi_{\pmb{\lambda}}(N')}{\sum_N \tilde\Psi_{\pmb{\lambda}}(N)\Psi_{\pmb{\lambda}}(N)} \, .
\ee

For numerical purposes, the operator with the elements $\hat{L}_{\pmb{\lambda}}(N,N')$ is truncated as a square matrix with boundaries $N_{\rm min}=N'_{\rm min}$ and $N_{\rm max}=N'_{\rm max}$. The solution of Eq.~(\ref{eq_modified_master_equation}) has the general form
\be
F(N,t)={\rm e}^{\hat{L}_{\pmb{\lambda}}t}F(N,0) \, ,
\ee
where the initial condition $F(N,0)$ can take arbitrary positive values. The matrix exponential ${\rm e}^{\hat{L}_{\pmb{\lambda}}t}$ can be computed using Pad\'e approximation. Since ${\rm e}^{\hat{L}_{\pmb{\lambda}}t}>0$,  the Perron-Frobenius theorem applies and the leading eigenvalue $-Q({\pmb{\lambda}})$ of $\hat{L}_{\pmb{\lambda}}$ corresponds to the the real maximum eigenvalue ${\rm e}^{-Q(\pmb{\lambda})t}$ of ${\rm e}^{\hat{L}_{\pmb{\lambda}}t}$ in magnitude (for some value $t>0$).  Consequently, the right-eigendistribution $\Psi_{\pmb{\lambda}}(N)$ can be asymptotically evaluated as
\be
\Psi_{\pmb{\lambda}}(N)\sim_{t\to\infty}{\rm e}^{\hat{L}_{\pmb{\lambda}}t}F(N,0) \, .
\ee

\begin{figure}[h]
\begin{minipage}[t]{0.55\hsize}
\resizebox{1.0\hsize}{!}{\includegraphics{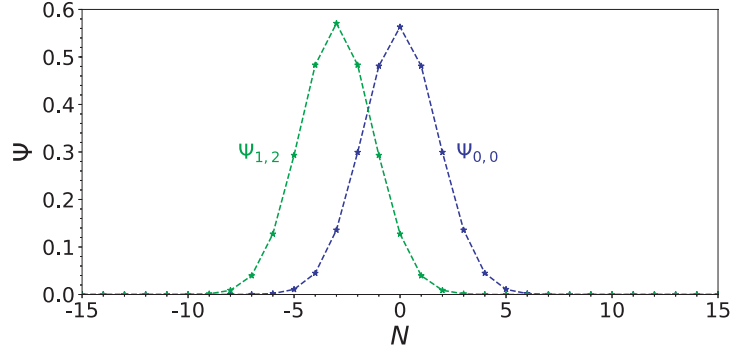}}
\end{minipage}
\caption{The normalized distribution $\Psi_{\lambda_1,\lambda_2}(N)$ for $\lambda_1=\lambda_2=0$ and for $\lambda_2=2\lambda_1=2$. Asterisks are from numerical results with dashed lines joining them. The parameters are $\beta=1$, $e=1$, $C_0=C_1=C_2=1$, $R_0=R_1=R_2=1$, $V_1=-1$, $V_2=-2$, and $V_p=0$.}
\label{fig_Phi}
\end{figure}

Figure \ref{fig_Phi} shows two examples of right-eigendistributions.  For $\pmb{\lambda}={\bf 0}$, the right-eigendistribution gives the stationary distribution as $P_{\rm st}(N)\sim \Psi_{\bf 0}(N)$.  The cumulant generating function $Q(\pmb{\lambda})$ can also be directly calculated by diagonalizing the matrix $\hat{L}_{\pmb{\lambda}}(N,N')$ and finding the largest eigenvalue.   The mean currents and higher cumulants can then be computed by numerically differentiating the so-obtained cumulant generating function with respect to the counting parameters~$\lambda_i$.

\subsection{Current-voltage characteristics}

\begin{figure}[h]
\begin{minipage}[t]{0.6\hsize}
\resizebox{1.0\hsize}{!}{\includegraphics{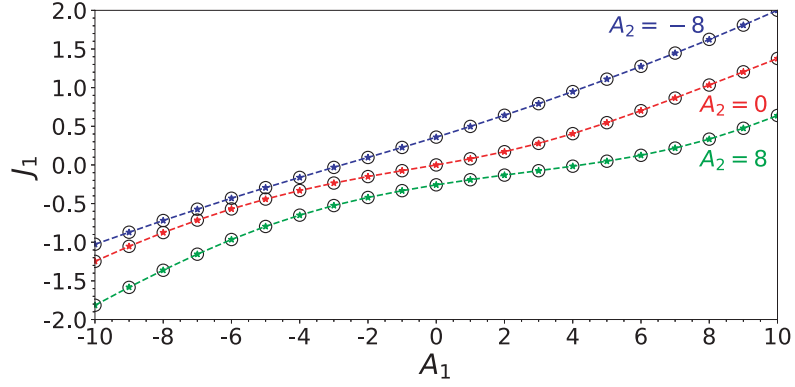}}
\end{minipage}
\caption{The mean electron current $J_1$ as a function of $A_1$, with $A_2$ fixed. The asterisks are the results from simulations and the circles are obtained by numerical differentiation of the cumulant generating function according to $J_1=(\partial Q/\partial\lambda_1)_{\pmb{\lambda}={\bf 0}}$. The dashed lines join the asterisks. The parameter values are $\beta=1$, $e=1$, $C_0=0.02$, $C_1=0.03$. $C_2=0.05$, $R_0=1$, $R_1=3$, $R_2=2$, and $V_p=3.0$.}
\label{fig_J_A}
\end{figure}

The mean electron current $J_1$ is depicted in Fig.~\ref{fig_J_A} as a function of the affinity $A_1$ for three different values of the other affinity $A_2$.  This current is obtained as in Eq.~(\ref{J}) from two different methods, by  numerical simulation (asterisks) and by differentiating the cumulant generating function with respect to $\lambda_1$ (circles).  The agreement between both methods is excellent.  The results show that the mean current has a nonlinear dependence on the affinity, so that its expansion around equilibrium should involve not only linear, but also nonlinear response coefficients.

\begin{figure}[h]
\begin{minipage}[t]{0.6\hsize}
\resizebox{1.0\hsize}{!}{\includegraphics{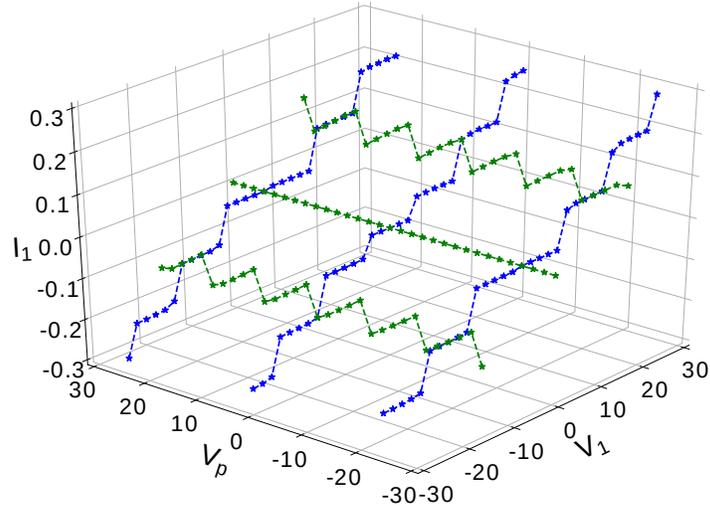}}
\end{minipage}
\caption{Current-voltage characteristics for the junction system when the parameters are chosen such that a pattern of Coulomb staircases is produced. The electric current defined as $I_1=-\vert e\vert J_1$ is shown versus the voltages $V_1$ and $V_p$.  The asterisks are numerical results from simulation and dashed lines join them. $I_1$-$V_1$ curves are depicted for $V_p=-24,0,24$ and $I_1$-$V_p$ curves for $V_1=-16,0,16$. The parameters are $\beta=100$, $e=1$, $C_1=0.1$, $C_0=C_2=0.001$, $R_1=100$, $R_0=R_2=1$, and $V_0=V_2=0$. This set of parameters is such that the junctions $0$ and $2$ play identical roles, so that an equivalent circuit would be given by replacing both junctions in parallel with a single one having a capacitance twice larger and a resistance half smaller, so that $I_2=-I_1/2$.}
\label{fig_I_V}
\end{figure}

The nonlinear dependence becomes stronger at lower temperature, as seen in Fig.~\ref{fig_I_V}  where the electric current~$I_1$ is shown as a function of the voltages $V_1$ and $V_p$, for $C_1\gg C_0=C_2$, $R_1\gg R_0=R_2$, and the temperature a hundred times lower than in Fig.~\ref{fig_J_A}.  We observe in Fig.~\ref{fig_I_V} that, under such conditions, Coulomb staircases manifest themselves in the current-voltage characteristics,  as for single-electron transistors.  The mean current $I_1$ as a function of the voltage $V_1$ forms plateaus separated by steep jumps.  Accordingly, the effective conductance $G=dI_1/dV_1$ forms sharp peaks, which is the manifestation of the Coulomb blockade effect, as in single-electron transistors \cite{K92}.  If $\beta\vert\Delta E_i^{(\pm)}\vert \gg 1$, the rates~(\ref{rates_model3}) are given by $W_i^{(\pm)}\simeq 0$ for $\Delta E_i^{(\pm)}>0$ and $W_i^{(\pm)}\simeq \vert\Delta E_i^{(\pm)}\vert/(e^2R_i)$ for $\Delta E_i^{(\pm)}<0$.  Therefore, the rates change around $\Delta E_i^{(\pm)}=0$ for low enough temperature.  Since the resistance $R_1$ is significantly larger than $R_0=R_2$, the rates $W_1^{(\pm)}$ play negligible roles with respect to $W_0^{(\pm)}$ and $W_2^{(\pm)}$.   Accordingly, the Coulomb staircases happen for $\Delta E_0^{(\pm)}=0$ and $\Delta E_2^{(\pm)}=0$.  For the parameter values taken in Fig.~\ref{fig_I_V} where $C_0=C_2$, $R_0=R_2$, and $V_0=V_2=0$, both conditions coincide, so that the Coulomb staircases appear on the lines given by
\be
V_1 = \frac{C_0+C_1+C_2}{C_1} \, V_p + \frac{e}{C_1} \left( N\mp \frac{1}{2} \right) \qquad \mbox{with} \quad N\in{\mathbb Z} 
\ee
in the plane $(V_p,V_1)$, as indeed observed in Fig.~\ref{fig_I_V}.  If $V_1=0$, the staircases are separated by $\Delta V_p=e/(C_0+C_1+C_2)$, corresponding to the period of current-voltage characteristics as a function of $V_p$.

\subsection{Graph analysis and affinities}

According to Schnakenberg's graph analysis \cite{S78}, the affinities driving the circuit out of equilibrium are given by
\bea
&& A_{1}=\ln\frac{W_1^{(+)}(N)\, W_0^{(-)}(N+1)}{W_0^{(+)}(N)\, W_1^{(-)}(N+1)}=-\beta eV_1 \, , \label{A1_model3}\\
&& A_{2}=\ln\frac{W_2^{(+)}(N)\, W_0^{(-)}(N+1)}{W_0^{(+)}(N)\, W_2^{(-)}(N+1)}=-\beta eV_2 \, . \label{A2_model3}
\eea

\begin{table}[h]
\caption{Numerical results for the nonlinear model of transistor. The numerical and theoretical affinities are compared for different conditions in the reservoirs. The voltages in the reservoirs are denoted $V_0$, $V_1$, and $V_2$. The parameter values are $\beta=1$, $e=1$, $C_0=C_1=C_2=1$, $R_0=R_1=R_2=1$, and $V_p=0$. The numerical affinities are computed over the time interval $t=10000$ with $1\times 10^4$ iterates.}
\vskip 0.3 cm
\begin{tabular}{rrrrrrr}
\hline
\hline
$\qquad V_0$ & $\qquad V_1$ & $\qquad V_2$ & $\qquad A_{10}^{\rm (th)}$ & $\qquad A_{10}^{\rm (num)}\qquad$ & $\qquad A_{20}^{\rm (th)}$ & $\qquad A_{20}^{\rm (num)}\qquad$ \bigstrut \\ \hline
$0.0$ & $0.7$ & $0.7$ & $-0.700$ & $-0.753 \pm 0.022$ & $-0.700$ & $-0.754 \pm 0.022$ \bigstrut \\ 
$0.0$ & $0.3$ & $0.7$ & $-0.300$ & $-0.330 \pm 0.014$ & $-0.700$ & $-0.737 \pm 0.017$ \bigstrut \\ 
$0.0$ & $0.3$ & $0.0$ & $-0.300$ & $-0.312 \pm 0.006$ & $0.000$ & $-0.008 \pm 0.004$ \bigstrut \\ 
$0.0$ & $0.3$ & $-0.3$ & $-0.300$ & $-0.311 \pm 0.007$ & $0.300$ & $0.304 \pm 0.007$ \bigstrut \\ 
$0.0$ & $-0.3$ & $0.5$ & $0.300$ & $0.300 \pm 0.009$ & $-0.500$ & $-0.504 \pm 0.011$ \bigstrut \\ 
$0.0$ & $-0.1$ & $0.1$ & $0.100$ & $0.103 \pm 0.002$ & $-0.100$ & $-0.100 \pm 0.002$ \bigstrut \\ 
$0.0$ & $-0.5$ & $0.6$ & $0.500$ & $0.518 \pm 0.013$ & $-0.600$ & $-0.615 \pm 0.014$ \bigstrut \\ 
$0.0$ & $-0.6$ & $-0.6$ & $0.600$ & $0.627 \pm 0.017$ & $0.600$ & $0.628 \pm 0.017$ \bigstrut \\ 
\hline
\hline
\end{tabular}
\label{tab_a_2}
\end{table}

In the same way as for the model with linear rates in Section~\ref{Model2}, the counting statistics of electron tunneling is carried out by simulation in order to compute numerically the affinities of this junction system. Since the system has a nonlinear character, we use the exact Gillespie algorithm. The mean values of the currents and diffusivities are evaluated by simulation and effective values are calculated for the rates by solving Eqs.~(\ref{W-J1})-(\ref{WWW}).  If the process was Poissonian, the affinities would thus be given by Eqs.~(\ref{A10})-(\ref{A20}).  The comparison between the so-obtained values of the affinities and the exact values are presented in Table~\ref{tab_a_2}. The values of the Poissonian hypothesis are relatively close to the exact values, in particular, in the vicinity of equilibrium ($A_1=A_2=0$), but discrepancies increase as the system is driven away from equilibrium.  The reason is that the rates of the model depend nonlinearly on the internal random variable $N$, so that the process is no longer Poissonian as it was the case for the models with constant and linear rates.  Neverthless, the approximation of the affinities provided by the Poissonian hypothesis is better than for the Gaussian approximation based on the central limit theorem.

\subsection{Fluctuation relation}

The symmetry of the fluctuation relation holds for the nonlinear model in the long-time limit.
Indeed, the operator defined by Eq.~(\ref{eq_modified_master_equation}) has the symmetry
\be\label{sym_rel_L}
\hat M^{-1} \, \hat{L}_{\pmb{\lambda}} \, \hat M = \hat{L}_{{\bf A}-\pmb{\lambda}}^{\dagger} 
\ee
with the affinities ${\bf A}=(A_1,A_2)$ and the operator $\hat M$ defined as the diagonal matrix $\hat M(N,N') = M(N) \, \delta_{N,N'}$ with the diagonal elements obeying the following recurrence,
\be\label{M-dfn}
M(N) = {\rm e}^{\beta \Delta E_0^{(-)}(N)} M(N-1) \, .
\ee
The symmetry~(\ref{sym_rel_L}) can be established by first noting that we should have $\Gamma_{\pmb{\lambda}}^{(-)}(N)M(N)=\Gamma_{{\bf A}-\pmb{\lambda}}^{(+)}(N-1)M(N-1)$. Next, the relations~(\ref{A1_model3}) and~(\ref{A2_model3}) for the affinities lead to the simplification $W_0^{(-)}(N)M(N)=W_0^{(+)}(N-1)M(N-1)$.  Finally, we have that $\Delta E_0^{(-)}(N)=-\Delta E_0^{(+)}(N-1)$ because of Eq.~(\ref{Delta_E}), so that the expression~(\ref{rates_model3}) for the rates implies that $W_0^{(+)}(N-1)/W_0^{(-)}(N)=\exp\left[\beta\Delta E_0^{(-)}(N)\right]$, hence the result~(\ref{M-dfn}).

As a consequence of the symmetry~(\ref{sym_rel_L}), it is known that its leading eigenvalue giving the cumulant generating function obeys the symmetry relation
\be\label{QFR_model3}
Q(\lambda_1,\lambda_2)=Q(A_1-\lambda_1,A_2-\lambda_2) \, ,
\ee
which is the direct consequence of asymptotic fluctuation relation~(\ref{eq_fluctuation_relation}) \cite{LM09,BEG11}.  

\begin{figure}[h]
\begin{minipage}[t]{0.5\hsize}
\resizebox{1.0\hsize}{!}{\includegraphics{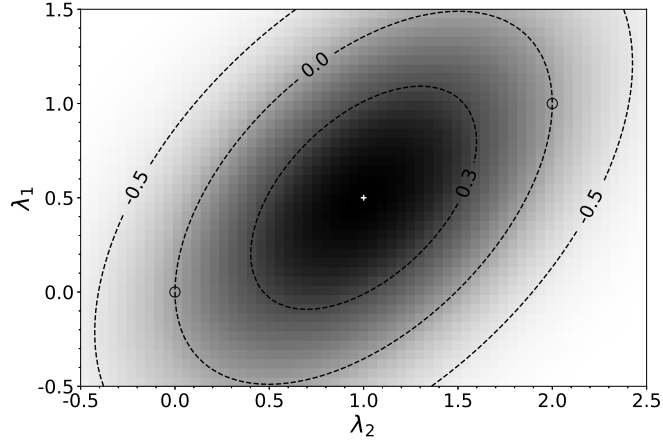}}
\end{minipage}
\caption{The gray scale map of the cumulant generating function $Q(\lambda_1,\lambda_2)$ computed by solving the eigenvalue problem~(\ref{eigenvalue}). Three contours are shown. The symbol plus marks the center located at $(\lambda_1=0.5,\lambda_2=1.0)$. The two circles indicate the coordinates $(\lambda_1=0.0,\lambda_2=0.0)$ and $(\lambda_1=1.0,\lambda_2=2.0)$, which are respectively joined by the contour for $Q(\lambda_1,\lambda_2)=0.0$. The parameter values are $\beta=1$, $e=1$, $C_0=C_1=C_2=1$, $R_0=R_1=R_2=1$, $V_1=-1$, $V_2=-2$, and $V_p=0$.}
\label{fig_map_Q}
\end{figure}

\begin{figure}[h]
\begin{minipage}[t]{0.99\hsize}
\resizebox{1.0\hsize}{!}{\includegraphics{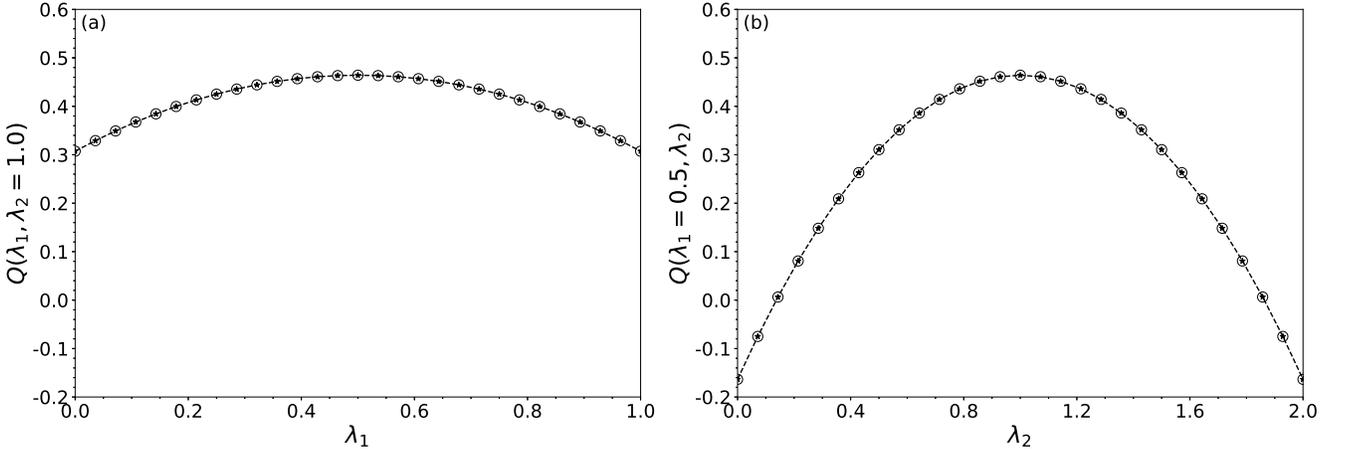}}
\end{minipage}
\caption{Two slices of the cumulant generating function $Q(\lambda_1,\lambda_2)$ computed by solving the eigenvalue problem~(\ref{eigenvalue}): (a)~the function versus $\lambda_1$ for $\lambda_2=1.0$; (b)~the function versus $\lambda_2$ for $\lambda_1=0.5$. In both panels, the asterisks are numerical results and the dashed lines join them. The circles are plotted from the function $Q(1.0-\lambda_1,\lambda_2=1.0)$ in panel (a) and $Q(\lambda_1=0.5,2.0-\lambda_2)$ in panel (b). Clearly, each circle surrounds an asterisk. The difference between each asterisk and the center of its corresponding circle is smaller than $1.0\times 10^{-14}$, so that the numerical error of cumulant generating function is negligible. The parameter values are $\beta=1$, $e=1$, $C_0=C_1=C_2=1$, $R_0=R_1=R_2=1$, $V_1=-1$, $V_2=-2$, and $V_p=0$.  The corresponding affinities are $A_1=1$ and $A_2=2$.}
\label{fig_Q}
\end{figure}

The symmetry relation~(\ref{QFR_model3}) is confirmed by the numerical evaluation of the cumulant generating function with the method explained in Subsection~\ref{Countstat_model3}.  In Fig.~\ref{fig_map_Q}, the density and several contour lines of the cumulant generating function $Q(\lambda_1,\lambda_2)$ are depicted in the plane of the counting parameters $\lambda_1$ and $\lambda_2$ for the affinities $A_1=1$ and $A_2=2$.  This figure shows the symmetry~(\ref{QFR_model3}) with respect to the inversion transformation $(\lambda_1,\lambda_2)\to(A_1-\lambda_1,A_2-\lambda_2)$.  We also observe that the cumulant generating function is not symmetric under the reflections $(\lambda_1,\lambda_2)\to(A_1-\lambda_1,\lambda_2)$ and $(\lambda_1,\lambda_2)\to(\lambda_1,A_2-\lambda_2)$, which is the evidence of coupling between the fluctuations of the two currents.

Moreover, Fig.~\ref{fig_Q} shows the cumulant generating function of Fig.~\ref{fig_map_Q} along the line $\lambda_2=1.0$ in panel~(a) and the line $\lambda_1=0.5$ in panel~(b).  These functions are compared with  their transformation by the symmetry~(\ref{QFR_model3}).  Their coincidence again confirms the validity of the fluctuation relation.

\subsection{Time-reversal symmetry relations for the response properties}

Here, we investigate the consequences of the fluctuation relation at the level of the linear and nonlinear response coefficients~(\ref{L_ij}) and~(\ref{M_ijk}).

Table~\ref{tab_relations_1} presents the values of all the linear response coefficients and diffusivities in the same conditions as in Fig.~\ref{fig_J_A}.  The fluctuation-dissipation relations (\ref{eq_FD_relation}) and the Onsager reciprocal relations (\ref{eq_Onsager_reciprocal_relations}) are verified up to numerical accuracy.

Beyond, Table~\ref{tab_relations_2} gives the values of the nonlinear response coefficients~(\ref{M_ijk}) and the first responses~(\ref{R_ijk}) of the diffusivities, again in the same conditions as in Fig.~\ref{fig_J_A}.  These values are obtained by numerical differentiations of the cumulant generating function computed as explained in Subsection~\ref{Countstat_model3}.  Table~\ref{tab_relations_2} shows that the relations~(\ref{eq_nonlinear_response_relations}) are also satisfied up to numerical accuracy.  Therefore, the implications of microreversibility beyond the linear regime are verified in this nonlinear model, as well as for the other models with constant and linear rates.

It should be mentioned that the response coefficients of even orders are all vanishing if $V_p$ is an integer multiple of $e/(C_0+C_1+C_2)$, because the transport properties have the symmetries $J_{i}({\bf A})+J_{i}({-\bf A})=0$ and $D_{ij}({\bf A})=D_{ij}({-\bf A})$ in this case.

\begin{table}[h]
\caption{The numerical values used in the fluctuation-dissipation relations (\ref{eq_FD_relation}) and the Onsager reciprocal relations (\ref{eq_Onsager_reciprocal_relations}). Three significant digits are obtained. The derivatives are approximated by numerical differentiations, which are accurate up to $O(10^{-4})$ with the method implemented here. The parameter values are the same as those used for Fig.~\ref{fig_J_A}.}
\begin{tabular}{>{\centering\arraybackslash}m{2cm}>{\centering\arraybackslash}m{3cm}>{\centering\arraybackslash}m{3cm}>{\centering\arraybackslash}m{4cm}}
\hline
\hline
$(i,j)$ & $L_{i,j}$ & $D_{ij}({\bf A}={\bf 0})$ & $L_{i,j}-D_{ij}({\bf A}={\bf 0})$ \bigstrut \\ \hline
$(1,1)$ & $7.59\times 10^{-2}$ & $7.58\times 10^{-2}$ & $1.15\times 10^{-4}$ \bigstrut \\
$(1,2)$ & $-2.53\times 10^{-2}$ & $-2.53\times 10^{-2}$ & $2.75\times 10^{-5}$ \bigstrut \\
$(2,1)$ & $-2.53\times 10^{-2}$ & $-2.53\times 10^{-2}$ & $3.03\times 10^{-5}$ \bigstrut \\
$(2,2)$ & $1.01\times 10^{-1}$ & $1.01\times 10^{-1}$ & $1.43\times 10^{-4}$ \bigstrut \\ \hline
\hline
\end{tabular}
\label{tab_relations_1}
\end{table}

\begin{table}[h]
\caption{The numerical values of the quantities used in the nonlinear response relations (\ref{eq_nonlinear_response_relations}). Three significant digits are obtained. The derivatives are approximated by numerical differentiations, which are accurate up to $O(10^{-6})$ with the method implemented here. The parameter values are the same as those used for Fig.~\ref{fig_J_A}.}
\begin{tabular}{>{\centering\arraybackslash}m{2.0cm}>{\centering\arraybackslash}m{3.0cm}>{\centering\arraybackslash}m{3.0cm}>{\centering\arraybackslash}m{3.0cm}>{\centering\arraybackslash}m{4.0cm}}
\hline
\hline
$(i,j,k)$ & $M_{i,jk}$ & $R_{ij,k}$ & $R_{ik,j}$ & $M_{i,jk}-R_{ij,k}-R_{ik,j}$  \bigstrut \\ \hline
$(1,1,1)$ & $5.27\times 10^{-3}$ & $2.63\times 10^{-3}$ & $2.63\times 10^{-3}$ & $3.60\times 10^{-6}$\bigstrut \\
$(1,2,2)$ & $2.54\times 10^{-3}$ & $1.27\times 10^{-3}$ & $1.27\times 10^{-3}$ & $-1.51\times 10^{-6}$ \bigstrut \\
$(1,1,2)$ & $-1.24\times 10^{-2}$ & $-1.15\times 10^{-2}$ & $-8.81\times 10^{-4}$ & $-7.48\times 10^{-6}$ \bigstrut \\
$(2,1,1)$ & $-1.76\times 10^{-3}$ & $-8.81\times 10^{-4}$ & $-8.81\times 10^{-4}$ & $3.26\times 10^{-6}$ \bigstrut \\
$(2,2,2)$ & $-1.01\times 10^{-2}$ & $-5.07\times 10^{-3}$ & $-5.07\times 10^{-3}$ & $-2.13\times 10^{-6}$ \bigstrut \\
$(2,1,2)$ & $-6.75\times 10^{-3}$ & $1.27\times 10^{-3}$ & $-8.01\times 10^{-3}$ & $-6.33\times 10^{-6}$ \bigstrut \\ \hline
\hline
\end{tabular}
\label{tab_relations_2}
\end{table}

\section{Conclusion and Perspectives}
\label{Conclusion}

In this paper, multivariate fluctuation relations have been investigated for the two currents coupled together in transistors, showing that their linear and nonlinear transport properties obey the Onsager reciprocal relations as well as their generalizations beyond the linear regime.  Multivariate fluctuation relations are symmetry relations finding their origin in microreversibility and expressed in terms of the thermodynamic forces or affinities driving the system out of equilibrium.  For transistors, these affinities are the two independent voltages that can be applied between the three ports.  The vehicles of our study are three stochastic models of transistors.  

For the first and second models with respectively constant and linear rates, the stochastic process is Poissonian allowing us to establish finite-time multivariate fluctuation relations in addition to the asymptotic multivariate fluctuation relations obtained in the long-time limit.  These finite-time multivariate fluctuation relations hold with respect to affinities (or voltages) that are time independent in the first model with constant rates, but for time-dependent affinities in the second model with linear rates.  These time-dependent affinities slowly converge as $1/t$ towards the values fixed by the particle reservoirs over a time scale determined by the rate constants of the transitions discharging the transistor.  In the models with constant and linear rates, the asymptotic values of the affinities can be recovered from the mean values of the currents and their diffusivities.  The reason is that the affinities as well as the currents and the diffusivities are all determined by the mean transition rates of these stationary Poissonian processes, so that they are related to each other by the nonlinear equations~(\ref{W-J1})-(\ref{A20}).

In contrast, for the third model with nonlinear rates, the stochastic process no longer has a Poissonian stationary probability distribution for its internal random variable and the affinities (or voltages) can no longer be exactly recovered by solving  the nonlinear equations~(\ref{W-J1})-(\ref{A20}).  Nevertheless, they continue to provide good approximations.  In this third model, a multivariate fluctuation relation is obtained in the long-time limit by proving a symmetry relation for the evolution operator modified to include counting parameters.

In all the three models of transistors, the multivariate fluctuation relation for the two currents always holds in the long-time limit.   The Onsager reciprocal relations and the fluctuation-dissipation relations are satisfied between the linear response coefficients and the equilibrium diffusivities in the linear regime close to equilibrium.  Furthermore, the relations between the nonlinear response coefficients and the first responses of the diffusivities are also satisfied as predicted by microreversibility in the nonlinear regimes away from equilibrium.
Although theoretically predicted, these results have not yet been tested experimentally and we think that such tests can be performed at room temperature on semiconducting transistors.

\section*{Acknowledgements}

Financial support from the China Scholarship Council under the Grant No. 201606950037, the Universit\'e Libre de Bruxelles (ULB), and the Fonds de la Recherche Scientifique - FNRS under the Grant PDR T.0094.16 for the project "SYMSTATPHYS" is acknowledged.

\vskip 0.5 cm

Jiayin Gu -- https://orcid.org/0000-0002-9868-8186

Pierre Gaspard -- https://orcid.org/0000-0003-3804-2110

\appendix

\section{Independent affinity relations}
\label{app_affinity_relations}

We introduce a convenient procedure to find all the independent affinity relations for a $n$-reservoir system by making an analogy with a polygon having $n$ vertices. Each affinity relation
\be
A_{ij}+A_{jk}=A_{ik}
\ee
can be conceived as a triangle with vertices $(i)$, $(j)$ and $(k)$. Thus, the question has now been transformed into the problem of finding the independent triangles within the associated polygon. Here, by "independent" we mean that any triangle cannot be represented by other triangles through vector analysis. 

\begin{figure}[h]
\begin{minipage}[t]{0.3\hsize}
\resizebox{1.0\hsize}{!}{\includegraphics{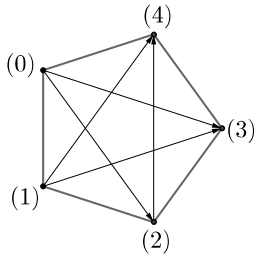}}
\end{minipage}
\caption{Polygon representing a system with $n$ reservoirs for calculating the total number of independent affinity relations. Here, the system is assumed to have $n=5$ reservoirs, so that the polygon is a pentagon in this example.}
\label{fig_pentagon}
\end{figure}

A pentagon corresponding to a $5$-reservoir system is taken as an example in Fig. \ref{fig_pentagon}. We describe the procedure as follows:
\begin{itemize}
\item Starting from the vertex $(0)$, sequentially draw ray lines to the vertices $(2)$, $(3)$, to the right-hand side of which we can find the triangles $\Delta_{(021)}$ and $\Delta_{(032)}$;
\item Starting from the vertex $(1)$, sequentially draw ray lines to the vertices $(3)$, $(4)$, to the right-hand side of which we can find the triangles $\Delta_{(132)}$ and $\Delta_{(143)}$;
\item Starting from the vertex $(2)$, there only exists one ray line to the vertex $(4)$, to the right-hand side of which we can find the triangle $\Delta_{(243)}$;
\item Combining the triangles found above and the pentagon itself, we can still find an extra independent triangle~$\Delta_{(034)}$.
\end{itemize}
Therefore, there is a total number of $6$ independent triangles, which equals the number of lines joining any two vertices within the pentagon plus an extra one. The reason for the extra one is evident when the polygon itself is a triangle. Generalizing to a polygon with $n$ vertices, we find that there exist $(n^2-3n+2)/2$
independent triangles, which are associated with the corresponding independent affinity relations.  Equation~(\ref{S1}) of Subsection~\ref{Countstat_model1} is thus obtained.

\section{Langevin stochastic process}
\label{app_Langevin_stochastic_process}

Introducing the probability density $\mathscr P$ in the limit where $N\gg 1$, the master equation~(\ref{eq_master_equation_10}) becomes
\be
\partial_t{\mathscr P}=\sum_{i=0}^2\biggl\{-\partial_N\Bigl[(W_{+i}-W_{-i}){\mathscr P}\Bigr]+\partial_N^2\left[\frac{1}{2}(W_{+i}+W_{-i}){\mathscr P}\right]\biggr\} \, ,
\ee
from which we see that the random variable $N$ obeys the following stochastic differential equation of Langevin type,
\be
\frac{{\rm d}N}{{\rm d}t}=\sum_{i=0}^2F_i
\ee
with the fluxes given by
\be
F_i=W_{+i}-W_{-i}+\sqrt{W_{+i}+W_{-i}}\, \xi_i(t)\, , \qquad\mbox{for}\qquad  i=0, 1, 2\, .
\ee
Here, $\xi_i(t)$ are Gaussian white noises satisfying the properties
\be
\langle\xi_i(t)\rangle=0\, ,\hspace{1cm}\langle\xi_i(t)\xi_j(t')\rangle=\delta_{ij}\, \delta(t-t') \, .
\ee
By discretization in time, we get
\be
N(t+\Delta t)=N(t)+\left(\sum_{i=0}^2F_i\right)\Delta t \, ,
\ee
with
\be
F_i=W_{+i}-W_{-i}+\sqrt{W_{+i}+W_{-i}}\, \frac{G_i(t)}{\sqrt{\Delta t}} \, , \qquad\mbox{for}\qquad  i=0,1,2 \, ,
\ee
where $G_i(t)$ are independent identically distributed normal random variables.

\end{document}